\documentclass[12pt,preprint]{aastex}

\slugcomment{Astrophys. J. in press}

\shorttitle{GRB-Jet Formation in Collapsars}
\shortauthors{Nagataki et al.}

\begin{document}


\title{Numerical Study on GRB-Jet Formation in Collapsars}


\author{Shigehiro Nagataki\altaffilmark{1,2}, Rohta
Takahashi\altaffilmark{3}, Akira Mizuta\altaffilmark{4}, 
Tomoya Takiwaki\altaffilmark{5}}

\altaffiltext{1}{Yukawa Institute for Theoretical Physics, Kyoto
University,
Oiwake-cho Kitashirakawa Sakyo-ku, Kyoto 606-8502, Japan, E-mail:
nagataki@yukawa.kyoto-u.ac.jp}
\altaffiltext{2}{KIPAC, Stanford University, P.O.Box 20450, MS 29,
Stanford, CA, 94309}
\altaffiltext{3}{Graduate School of Arts and Sciences,
The University of Tokyo, Tokyo 153-8902, Japan}
\altaffiltext{4}{Max-Planck-Institute f$\rm \ddot{u}$r Astrophysik,
Karl-Schwarzschild-Str.
1, 85741 Garching, Germany}
\altaffiltext{5}{Department of Physics, The University of Tokyo,
Bunkyo-ku, Tokyo 113-0033, Japan}

\begin{abstract}
Two-dimensional magnetohydrodynamic simulations are performed
using the ZEUS-2D code to investigate the dynamics of a collapsar
that generates a GRB jet, taking account of realistic equation of 
state, neutrino cooling and heating processes, magnetic fields,
and gravitational force from the central black hole and self gravity. 
It is found that neutrino heating processes are not so efficient to
launch a jet in this study. 
It is also found that a jet
is launched mainly by $B_\phi$ fields that are amplified by the
winding-up effect.
However, since the ratio of total energy relative to the rest
mass energy in the jet
is not so high as several hundred, we conclude that
the jets seen in this study
are not be a GRB jet. This result suggests that general relativistic effects,
which are not included in this study, will be important to generate a GRB jet. 
Also, the accretion disk with magnetic fields may still play an important
role to launch a GRB jet, although a simulation for much longer
physical time ($\sim 10-100$ s) is required to confirm this effect.
It is shown that considerable amount of $\rm ^{56}Ni$ is synthesized
in the accretion disk. Thus there will be a possibility
for the accretion disk to
supply sufficient amount of $\rm ^{56}Ni$ required to explain the
luminosity of a hypernova. 
Also, it is shown that neutron-rich matter due to electron captures with high 
entropy per baryon is ejected along the polar axis. Moreover,
it is found that the electron fraction becomes larger than 0.5 around the 
polar axis near the black hole by $\nu_e$ capture at the region.
Thus there will be a possibility that $r$-process and $r/p-$process nucleosynthesis
occur at these regions. Finally, much neutrons will be
ejected from the jet, which suggests that 
signals from the neutron decays may be observed as the delayed bump
of the light curve of the afterglow or gamma-rays. 
\end{abstract}

\section{INTRODUCTION}\label{intro}

There has been growing evidence linking long gamma-ray bursts (GRBs;
in this study, we consider only long GRBs, so we refer to long GRBs as
GRBs hereafter for simplicity) to the death of massive stars.
The host galaxies of GRBs are
star-forming galaxies and the positions of GRBs appear to trace the
blue light of young stars~\citep{vreeswijk01,bloom02,gorosabel03}.
Also, 'bumps' observed in some afterglows can be naturally explained
as contribution of bright
supernovae~\citep{bloom99,reichart99,galama00,garnavich03}.
Moreover, direct evidence of some GRBs accompanied by supernovae
have been reported such as the association of GRB 980425 with
SN 1998bw~\citep{galama98,iwamoto98}, that of GRB 030329 with SN
2003dh~\citep{hjorth03,price03,stanek03}, and that of GRB 060218
and SN 2006aj~\citep{mirabal06,mazzali06}.

It should be noted that these supernovae (except for SN 2006aj)
are categorized as a new
type of supernovae with large kinetic energy ($\sim 10^{52}$ ergs),
nickel mass ($\sim 0.5M_{\odot}$), and
luminosity~\citep{iwamoto98,woosley99}, so these supernovae are
sometimes called hypernovae. The total explosion energy of the
order of $10^{52}$ erg is too important to be emphasized, because
it is generally considered that a normal core-collapse supernova 
cannot cause such an energetic explosion. 
Thus another scenario has to be considered to explain the system of
a GRB associated with a hypernova. 
One of the most promising scenarios is the collapsar
scenario~\cite{woosley93}. In the collapsar scenario, a black hole
is formed as a result of gravitational collapse. Also, rotation of
the progenitor plays an essential role. Due to the rotation, an
accretion disk is formed around the equatorial
plane. On the other hand, the matter around the rotation axis freely
falls
into the black hole. MacFadyen and Woosley (1999) pointed out 
that the jet-induced explosion along the rotation axis may occur
due to the heating through neutrino anti-neutrino pair annihilation
that are emitted from the accretion disk (see also Fryer
and M$\rm \acute{e}$sz$\rm \acute{a}$ros 2000). 

It is true that the collapsar scenario is the breakthrough
on the problem of the central engine of GRBs. However, there are many
effects that have to be involved in order to establish the scenario
firmly. First of all, neutrino heating effects have to be
investigated carefully by including microphysics of neutrino processes
in the numerical simulations. It is true that MacFadyen and Woosley (1999) 
have done the numerical simulations of the collapsar, in which
a jet is launched along the rotation axis,
but detailed microphysics of neutrino heating is not included in their
simulations. Secondly, it was pointed out that effects
of magnetic fields and rotation may play an important role to launch
the GRB jets~\citep{proga03,mizuno04a,mizuno04b,proga05,shibata05,sekiguchi05,fujimoto06}, although 
neutrino heating effects are not included in their works. 
Recently, Rockefeller et al (2006) presented 3-dimensional simulations of collapsars 
with smoothed particle hydrodynamics code. In their study, 3-flavor flux-limited
diffusion package is used to take into account neutrino cooling and absorption
of electron-type neutrinos, although neutrino anti-neutrino pair annihilation
is not included. They have shown that alpha-viscosity drives energetic
explosion through 3-dimensional instabilities and angular momentum 
transfer, although the jet is not launched and magnetic fields (source of 
the viscosity) are not included in their study.
Thus it is not clear
which effects are most important to launch a GRB jet, that is, what
process is essential as the central engine of GRBs. 

Due to the motivation mentioned above, we have performed
two-dimensional magnetohydrodynamic simulations of collapsars with
magnetic fields, rotation, and neutrino cooling/heating processes.
In our simulations, the realistic equation of state (EOS) of Blinnikov
et al. (1996) and effects of photo-disintegration of nuclei are also
included. We investigated influence of magnetic fields on the dynamics
of collapsars by changing initial amplitude of the magnetic fields.
In section~\ref{method}, models and numerical methods are explained.
Results are shown in section~\ref{results}. Discussions are described
in~\ref{discussions}. Summary and conclusion are presented in
section~\ref{conclusion}.

\section{MODELS AND NUMERICAL METHODS}\label{method}

Our models and numerical methods of simulations in this study are
shown in this section. First we present equations of ideal MHD,
then initial and boundary conditions are explained. Micro physics
included in this study (equation of state (EOS), nuclear reactions,
and neutrino processes) is also explained.

\subsection{Magnetohydrodynamics}\label{hydro}

We have done two-dimensional MHD simulations
taking account of self-gravity and gravitational potential of
the central point mass. The calculated
region corresponds to a quarter of the meridian plane under the
assumption of axisymmetry and equatorial symmetry. 
The spherical mesh with 150($r$)$\times$ 30($\theta$) grid points
is used for all the computations. The radial grid is nonuniform,
extending from $r=$1.0$\times 10^{6}$ to 1.0$\times 10^{10}$ cm
with finer grids near the center, while the polar grid is uniform.
The minimum radial grid is set to be 3.0$\times 10^{5}$ cm.
We have confirmed that a free flow is smoothly solved with this grid
resolution.

The basic equations in the following form are finite differenced
on the spherical coordinates:
\begin{eqnarray}
\frac{D \rho}{D t} = && - \rho \nabla \cdot \bf{v} 
\label{Eq:basic0} \\
\rho \frac{D \bf{v}}{D t} = && - \nabla p - \rho \nabla (\Phi_p + \Phi_s) 
+ \frac{1}{4 \pi} (\nabla \times \bf{B}) \times \bf{B} \\
\rho \frac{D}{D t} \left(   \frac{e}{\rho}  \right) = && - p \nabla
\cdot \bf{v} - L_{\nu}^{-} + L_{\nu}^{+} + L_{\rm nucl}    \\
\frac{\partial \bf{B}}{\partial t} = && \nabla \times (\bf{v} \times
\bf{B}) \\
\Delta \Phi_s = && 4 \pi G \rho \\
\frac{D Y_e}{Dt} = && - Y_p \Gamma_{p \rightarrow n} + Y_n \Gamma_{n \rightarrow p},
\label{Eq:basic1}
\end{eqnarray}
where $\rho$, $\bf{v}$, $p$, $\Phi_p$, $\Phi_s$
$e$, $L_{\nu}^{\pm}$, $L_{\rm nucl}$, $Y_e$, $Y_p$, $Y_n$, $\Gamma_{{p \rightarrow n}}$, and $\Gamma_{{n \rightarrow p}}$, and 
are density, velocity, pressure, gravitational potential due to the
central point
mass (black hole), gravitational potential due to self gravity,
internal energy density, heating/cooling rates due to
neutrino processes, energy gain (loss) rate due to nuclear
reaction, fraction of electron, proton, neutron, and reaction rate
from proton to neutron (electron capture rate plus $\nu_e$ capture on neutron) 
and from neutron to proton (positron capture plus 
$\bar{\nu}_e$ capture on proton), respectively.
The Lagrangian derivative is denoted as $D/D t$. The gravitational
potential of the black hole is modified to take into account some of
the general relativistic effects~\cite{paczynski80}, $\Phi_p = - GM/(r
- r_{\rm s})$ where $r_{\rm s} = 2GM/c^2$ is the Schwartzshild radius.
Self-gravity is obtained in ZEUS-2D code developed by Stone and Norman
(1992a,b), by solving the matrix which results from finite-differencing
the Poisson equation in two dimension. The ZEUS-2D code is also used
to solve the MHD equations with second order accurate interpolation in
space. 
Energy gain (loss) rate due to nuclear reaction and heating/cooling rates 
due to neutrino processes are described in
subsections~\ref{nucl} and~\ref{neutrino}. Effects of
$\alpha$-viscosity and (anomalous) resistivity are not included in
this study to avoid the uncertainty of the treatment of these effects.

\subsection{Initial and Boundary Conditions}\label{initial}

We adopt the model E25 in Heger et al. (2000). This model corresponds
to a star that
has 25$M_{\odot}$ initially with solar metallicity, but loses
its mass and becomes to be 5.45$M_{\odot}$ of a Wolf-Rayet star at the
final stage. This model seems to be a good candidate as a progenitor of a
GRB since losing their envelope will be suitable to 
be a Type Ic-like supernova and to make a baryon poor fireball. 
The mass of the iron core is 1.69$M_{\odot}$ in this model.
Thus we assume that the iron core has collapsed and formed a black hole
at the center. This treatment is same with Proga et al. (2003).
The Schwartzshild radius of the black hole is 5.0$\times 10^5$ cm initially.

We explain how the angular momentum is distributed initially. 
At first, we performed
1-D simulation for the spherical collapse of the progenitor
for 0.1 s when the inner most Si-layer falls to the inner most 
boundary (=$10^6$ cm). After the spherical collapse,
angular momentum was distributed so as to provide a constant ratio of
0.05 of centrifugal force to the component of gravitational force
perpendicular to the rotation axis at all angles and radii, except
where that prescription resulted in $j_{16}$ greater than a prescribed
maximum value, 10. This treatment is similar to the one in MacFadyen and
Woosley (1999). The total initial rotation energy is $2.44 \times 10^{49}$
erg that corresponds to $1.3 \times 10^{-2}$ for initial ratio of 
the rotation energy to the gravitational energy ($T/W$).

Configuration and amplitude of the magnetic fields in a progenitor
prior to collapse are still uncertain. Thus in this study we choose
a simple form for the initial configuration prior to collapse
and the amplitude is
changed parametrically. Initial configuration of the magnetic fields
is chosen as follows:
\begin{eqnarray}
\vec{B}(\vec{r}) = && \frac{1}{3} B_0 \left(  \frac{r_0}{r}   \right)^3 (2
\cos \theta \vec{e_r}  + \sin \theta \vec{e_{\theta}} )  \;\;\; \rm
for \;\; r \ge r_0 \\
                 = && \frac{2}{3} B_0 (
\cos \theta \vec{e_r}  - \sin \theta \vec{e_{\theta}} )  \;\;\; \rm
for \;\; r < r_0.
\end{eqnarray}
This configuration represents that the magnetic fields are uniform in
a sphere $(r < r_0)$, while dipole at outside of the sphere. We set
$r_0$ to be the boundary between CO core/He layer (= $3.6 \times 10^9$
cm). $B_0$ corresponds to the strength of the magnetic field in the
sphere. We have chosen $B_0$ to be 0, $10^8$G, $10^9$G,
$10^{10}$G, $10^{11}$G, and $10^{12}$G.

As for the boundary condition in the radial direction,
we adopt the outflow boundary condition for the inner and outer
boundaries. That is, the flow from the central black hole is prohibited
at the inner boundary and the inflow from the surface of the
progenitor is prohibited at the outer boundary. Of course, the mass of
the central black hole becomes larger due to the mass accretion from
the inner boundary.
As for the boundary condition in the zenith angle direction, axis of
symmetry boundary condition is adopted for the rotation axis, while
the reflecting boundary condition is adopted for the equatorial plane.
As for the magnetic fields, the equatorial symmetry boundary condition,
in which the normal component is continuous and the tangential component 
is reflected, is adopted.

\subsection{Micro Physics}

\subsubsection{Equation of State}\label{eos}
The equation of state (EOS) used in this study is the one
developed by Blinnikov et al. (1996). This EOS contains
an electron-positron gas with arbitrary degeneracy, which is in thermal
equilibrium with blackbody radiation and ideal gas of
nuclei. We used the mean atomic weight of nuclei to estimate the
ideal gas contribution to the total pressure, although
its contribution is negligible relative to those of electron-positron 
gas and thermal radiation in our simulations.

\subsubsection{Nuclear Reactions}\label{nucl}
Although the ideal gas contribution of nuclei to the total pressure
is negligible, effects of energy gain/loss due to nuclear reactions
are important. In this study, nuclear statistical equilibrium (NSE)
was assumed for the region where $T \ge 5 \times 10^9$ [K] is
satisfied. This treatment is based on the assumption
that the timescale to reach and maintain NSE is much shorter than the
hydrodynamical time. Note that complete Si-burning occurs in explosive
nucleosynthesis of core-collapse supernovae for the region $T \ge 5
\times 10^9$ [K]~\cite{thielemann96}.
The hydrodynamical time in this study, $\sim
\rm s$ (as shown in Figs~\ref{fig3},~\ref{fig4}, and~\ref{fig10},
the nuclear reaction
occurs at $\sim (10^7 - 10^8)$cm 
where the radial velocity is of the order of 
$(10^7-10^8)$ cm s$^{-1}$), 
is comparable to the explosive nucleosynthesis in core-collapse
supernovae, so the assumption of NSE adopted in this study seems to be 
well. 5 nuclei, $\rm n,p,^4He,^{16}O$, and $\rm ^{56}Ni$ were
used to estimate the binding energy of ideal gas of nuclei in NSE for
given ($\rho$, $T$, $Y_e$). $Y_e$ is electron fraction that is
obtained from the calculations of neutrino processes in
section~\ref{neutrino}. On the other hand, we assumed that  
no nuclear reaction occurs for the region where
$ T < 5 \times 10^9$ [K].

\subsubsection{Neutrino Processes}\label{neutrino}

Neutrino cooling processes due to pair capture on free nucleons,
pair annihilation, and plasmon decay are included in this study. 
Since photoneutrino and bremsstrahlung processes are less important
ones at $10^{9} < T < 10^{11} $ [K] and $\rho < 10^{12} $ [g
cm$^{-3}$]~\cite{itoh89} where neutrino cooling effects are
important in our calculations, we do not include these processes.

As for the electron capture on free proton, we extend the formulation
of Epstein and Pethick (1981) to arbitrary degeneracy. 
The energy loss rate per proton $\dot{Q}_{\rm EC}$ [GeV s$^{-1}$ proton$^{-1}$ ] is 
\begin{eqnarray}
\dot{Q}_{\rm EC} =&& \frac{G_\beta^2 (C_V^2 +3G_A^2)}{2\pi^3\hbar } 
\left\{ (k_B
T)^6 F_5(x) + 2Q_v (k_B T)^5 F_4(x) + Q_v^2 (k_B T)^4 F_3(x) 
\right \} \\ \nonumber
&& \times \frac{\exp (F_5 (\eta) /F_4 (\eta)  ) }{1 + \exp (F_5 (\eta)
/F_4 (\eta)) },
\label{Eq:EC}
\end{eqnarray}
where $G_\beta = G_F \cos \theta_c$ (in units of GeV$^{-2}$)
is the Fermi coupling constant
($\theta_c$ is the Cabbibo angle), $C_V = 1/2 + 2 \sin ^2 \theta_W$ is
the vector coupling constant ($\theta_W$ is the Weinberg angle), 
$C_A = 1/2$ is the axial-vector coupling constant, $k_B$ is the
Boltzmann constant, and $Q_v = (m_n - m_p)c^2 $ is mass difference
between neutron and proton. $x$ and $\eta$ are defined as
$x = (\mu_e - Q_v)/(k_B T)$ and $\eta = \mu_e/(k_B T)$, where $\mu_e$
is the chemical potential of electrons. $F_i (x)$ is the Fermi
integral that is defined as
\begin{eqnarray}
F_i (x) = \int_0^\infty \frac{y^k}{ 1 + e^{y-x}} dy.
\end{eqnarray}
We used the formulation derived by Takahashi et al. (1978) to solve the
Fermi integral. Note that the leading term becomes the first term in 
Eq.~\ref{Eq:EC} and obeys $\propto T^6$ when electrons are not degenerated,
although other terms dominate when electrons become to be degenerated. 
We also include the Fermi blocking
factor~\cite{herant94}
due to $\nu_e$ assuming that the chemical potential of $\nu_e$ is
zero and mean
energy of the emitted neutrinos by this process is $k_B T F_5 (\eta) /F_4
(\eta)$~\cite{rosswog03}. All we have to do is to change
$\mu_e$ to -$\mu_e$ and $Q_v$ to -$Q_v$ to estimate the
positron capture on free neutron.
As for the plasmon decays, we used the formulation of Itoh et
al. (1989) with the blocking factor~\cite{herant94} assuming that
the chemical potential of $\nu_e$ is zero and mean
energy of the emitted neutrinos is $k_B T \left \{ 1 + \gamma_{\rm
loss}^2/2(1+ \gamma_{\rm loss}) \right \}$ where $\gamma_{\rm loss} = 5.565
\times 10^{-2}\sqrt{\pi^2/3 + \eta^2} $~\cite{rosswog03}.
We have also used the formulation of Itoh et al. (1999) for pair
annihilation with the blocking factor~\cite{herant94}. The mean energies
of $\nu_e$ and $\bar{\nu}_e$ emitted by this process
are $k_B T F_4(\eta)/F_3(\eta)$ and $k_B
T F_4(- \eta)/F_3(- \eta)$, respectively~\cite{cooperstein86}. 

In our simulations, density at the inner most region reaches
as high as $10^{12}$ g cm$^{-3}$ where nuclei are almost 
photo-disintegrated into nucleons. In this region, as discussed
in section~\ref{neutrino-dis}, optically thin limit breaks down. 
In order to take into account such a high density region, neutrino
leakage scheme is introduced~\cite{kotake03}. In this scheme,
we calculate typical diffusion timescale for neutrinos to escape 
from the system. Since chemical composition is dominated by nucleons
at such a high density region, the opacity is mainly determined
by scattering and absorption by nucleons for $\nu_e$ and $\bar{\nu}_e$
and scattering by nucleons for $\nu_\mu$, $\bar{\nu}_\mu$, 
$\nu_\tau$, and $\bar{\nu}_\tau$. The mean free path can be written
as 
\begin{eqnarray}
\lambda_{\nu_e} ^{-1}&& = \frac{\rho}{m_u} \sigma_{\rm sc}(\epsilon_{\nu_e}) 
+ \frac{\rho Y_n}{m_u} \sigma_{\rm ab}(\epsilon_{\nu_e})   
\;\;\; \rm for \;\; \nu_e \\  
\lambda_{\bar{\nu}_e}^{-1}&& = \frac{\rho}{m_u} \sigma_{\rm sc} 
(\epsilon_{\bar{\nu}_e}) 
+ \frac{\rho Y_p}{m_u} \sigma_{\rm ab} (\epsilon_{\bar{\nu}_e}) 
 \;\;\; \rm for \;\; \bar{\nu}_e   \\
\lambda_{\nu_x}^{-1}&& = \frac{\rho}{m_u} \sigma_{\rm sc}(\epsilon_{\nu_x})  
\;\;\; \rm for \;\; other \;\; flavors
\end{eqnarray}
where $m_u$, $\sigma_{\rm sc}$, $\sigma_{\rm ab}$, $Y_p$, $Y_n$,
$\epsilon_{\nu_e}$, $\epsilon_{\bar{\nu}_e}$, and $\epsilon_{\nu_x}$
are the atomic mass unit, scattering cross section by nucleon,
absorption cross section by nucleon, mass fraction of proton and neutron,
and mean energies of $\nu_e$, $\bar{\nu}_e$, and $\nu_x$.
We take the cross sections for these interactions from Bethe (1990).
The diffusion timescale is given by
\begin{eqnarray}
\tau_{\rm diff} = \frac{3 \Delta R^2}{\pi^2 c \lambda}
\end{eqnarray}
where we take $\Delta R = (r \Delta r \Delta \theta)^{1/2}$
as the typical diffusion length.
$\Delta r$ and $r \Delta \theta$ are grid sizes in the radial and
polar directions at that point. Neutrino emissivity [GeV cm$^{-3}$ s$^{-1}$] 
at the point is calculated as
\begin{eqnarray}
f_{\nu}^{\rm leak} = f_{\nu} \times \rm min \left(
1.0, \frac{dt}{\tau_{\rm diff}}
\right ) 
\end{eqnarray}
where $f_{\nu}$ and $dt$ are neutrino emissivity 
obtained by the optically thin limit and time step of the simulations.
On the other hand, when we estimate the neutrino heating rate, we take
optically thin limit for all the regions for simplicity. 
This corresponds to that we estimate the upper limit for the
neutrino heating effects. Even this limit, as shown later, jet like explosion
does not occur by the neutrino heating (as shown in Fig.~\ref{fig1}). 
Also, this optically thin limit will be  justified since neutrino 
cooling rate dominates heating rate, as shown in Fig.~\ref{fig8}.
Detailed discussion on this point is presented in section~\ref{neutrino-dis}.

Neutrino heating process due to neutrino pair annihilation and 
$\nu_e$ and $\bar{\nu}_e$ captures on free nucleons with blocking factors of
electrons and positrons are included in this study. As for the
neutrino pair annihilation process, the formulation of Goodman et
al. (1987) (with blocking factors) is adopted. The $\nu_e$ and
$\bar{\nu}_e$ captures on free nucleons are inverse processes of
electron/positron captures. The calculation of neutrino heating is the most
expensive in the simulation of this study. Thus, to save the CPU time, 
the neutrino heating processes are calculated only within the limited 
regions ($r < 10^9$ cm). Moreover, we adopt some criterions as
follows in order to save CPU time.
We have to determine the energy deposition regions and emission regions 
to estimate the neutrino heating rate. 
As for the neutrino pair annihilation process,
We adopt no criterion for the energy deposition regions other than $r < 10^9$
cm, while the emission regions satisfy the criterion of
$T \ge 3 \times 10^9$ K. 
As for the neutrino capture process, we adopt the criterions as follows:
The absorption regions should satisfy the criterion of $\rho \ge 10^4$ g
cm$^{-3}$, while the emission regions satisfy the criterion of (i)
$\rho \ge 10^3$ g cm$^{-3}$ and (ii) $T \ge 10^9$ K. 
Also, these heating rate had to be updated every 100 time steps to
save CPU time. Of course this treatment has to be improved in the future.
However, this treatment seems to be justified as follows: The total time step 
is of the order of $10^6$ steps. The final physical time is of the
order of seconds, so the typical time step is $\sim 10^{-6}$ s. This
means that heating rate is updated every $\sim 10^{-4}$ s, which
will be shorter than the typical dynamical timescale. 
This point is discussed in detail in section~\ref{neutrino-dis}

The heating process of neutrino pair
annihilation are calculated as follows. 
Let points A and B be in the emission regions
and a point C be in the energy deposition regions. Let $d_A$, $d_B$,
and $d_C$ be distances between A and B, B and C, and C and A,
respectively. The angle $\psi$ is defined as the angle ACB. The energy
deposition rate at C, $\dot{Q}(C)$ [GeV cm$^{-3}$ s$^{-1}$],
is given formally by integrating the emission regions as
\begin{eqnarray}
\dot{Q}(C) = \frac{K G_F^2}{2 \pi c} \int \int dx^3_A dx^3_B \frac{(1
- \cos\psi)^2 }{d_A^2 d_B^2} \int d \epsilon_{\nu} d
  \epsilon_{\bar{\nu}} \epsilon_{\nu} \epsilon_{\bar{\nu}}
  (\epsilon_{\nu} + \epsilon_{\bar{\nu}}) F_{\nu} F_{\bar{\nu}} (1 -
  f_e^-) (1 - f_e^+),
\label{Eq:anni} 
\end{eqnarray}
where $K$ is $(1 - 4 \sin^2 \theta_W + 8 \sin^4 \theta_W) /(6 \pi)$
for $\nu_{\mu} \nu_{\bar{\mu}}$ and $\nu_{\tau} \nu_{\bar{\tau}}$
annihilation and $(1 + 4 \sin^2 \theta_W + 8 \sin^4 \theta_W) /(6 \pi)$
for $\nu_{e} \nu_{\bar{e}}$ annihilation~\cite{goodman87}, $\bf{x_A}$
and $\bf{x_B}$ are locations of the points A and B, $\epsilon_{\nu,
\bar{\nu}}$ are energies of (anti-)neutrinos, $F_{\nu, \bar{\nu}}$ are
energy spectrum [cm$^{-3}$ s$^{-1}$ GeV$^{-1}$] of (anti-)neutrinos,
and $f_e^{\pm}$ are Fermi blocking factors of electrons/positrons at
point C. In this formulation, $G_F$ is written in units of cm GeV$^{-1}$.
It is easily seen that Eq.~(\ref{Eq:anni}) is eight dimensional
integral and it takes too much time to accomplish this integral. 
Thus we have approximated that each neutrino energy spectrum due to each
emission process is a monotonic one, that is, we assumed 
only neutrinos with average energy are emitted when Eq.~(\ref{Eq:anni})
is carried out. Since the energy loss rate [GeV cm$^{-3}$
s$^{-1}$] is obtained from the formulations of neutrino cooling
process mentioned above, number of emitted neutrinos for each process
is calculated by dividing the average energy of emitted neutrinos for
the process.
Although the calculated region is a quarter of the meridian plane, the
neutrino flux from the south hemisphere is included assuming the
symmetry relative to the equatorial plane.
The same
treatment with neutrino pair annihilation process is adopted to
estimate the heating rate due to $\nu_e$ and $\bar{\nu}_e$ capture
processes, that is, average energy
of (anti-)electron-type neutrinos is used.

Neutrinos are emitted isotropically in the fluid-rest frame, so strictly
speaking, neutrinos are emitted unisotropically in the coordinate system
due to the beaming effect~\cite{rybicki79}. In fact, the angular frequency
is found to become as large as $10^4$ s$^{-1}$ at the inner
most region (Figs.~\ref{fig3} and~\ref{fig10}) and the rotation
velocity, $v_\phi$ reaches $\sim 10^{10}$ cm s$^{-1}$ at most.
Thus the beaming effect may be important although we did not
take into account the effect in this study.

Finally the models considered in this study are summarized in
Table~\ref{tab1}. The digit in the name of each model represents
the power index of $B_0$. 

\placetable{tab1}

\section{RESULTS}\label{results}

We present results of numerical simulations in this section.
First of all, we show in Fig~\ref{fig1} the density contour of the
central region of the progenitor ($r \le 10^8$cm) with
velocity fields for model 0 and model 9.
In the top panels, density contour with velocity fields for model 0
at $t$ = 2.1 s (left panel) and $t$ = 2.2 s (right panel) is shown,
while in the bottom panels the ones for model 9 is shown.
The color represents the density (g cm$^{-3}$) in logarithmic scale
($10^{3}-10^{12}$). Vertical axis and horizontal axis represent
polar axis (= rotation axis) and equatorial plane, respectively.
You can easily find
that a jet is launched at $t$ = 2.1 s for model 9, while no jet
occurs for model 0. In the following subsections, we explain the
dynamics of these models in detail.

\placefigure{fig1}

\subsection{Dynamics without Magnetic Fields}\label{withoutmag}

In this subsection, we mainly explain the dynamics of model 0, then
the one of model 9 is explained in the next subsection.
First, we show the accreted energy and mass
accretion rate as a function of time in Fig.~\ref{fig2}(a).
Solid, short-dashed, and dotted lines represent
accreted total energy, kinetic energy, and thermal energy
($\times 10^{50}$ erg) into the black hole. 
Long-dashed line represents the mass accretion rate ($M_{\odot}$
s$^{-1}$) as a function of time. From this figure, we can understand
the following points:
(i) the accreted energy amounts to of the order of $10^{52}$ erg,
which is comparable to the explosion energy of a hypernova. Thus there
is a possibility for the models in this study to explain the large
explosion energy of a hypernova as long as the released gravitational energy
is efficiently converted to the explosion energy.
(ii) kinetic energy dominates thermal energy at the inner
boundary. This is because almost all thermal energy is extracted in
the form of neutrinos, which is seen in Fig.~\ref{fig6}(a). 
(iii) mass accretion rate drops almost monotonically from $\sim
10^{-1}M_{\odot}$ s$^{-1}$ to $\sim 10^{-3}M_{\odot}$ s$^{-1}$, which is
discussed in sections~\ref{withmag} and~\ref{b-fields-dis}.

\placefigure{fig2}

Next, we show in Fig.~\ref{fig3} profiles of physical quanta of the 
accretion disk around the equatorial plane.
Profiles of density, absolute value of
radial velocity, angular frequency, temperature, density scale height,
and specific angular momentum in the accretion disk are shown for model 0 at
$t$ = 2.2 s. From the figure, we can understand the following
points:
(i) the density reaches as high as $10^{12}$ g cm$^{-3}$ around the
central region. 
(ii) the inflow velocity ($v_r$) becomes as low as 10$^6$ cm s$^{-1}$ at the
central region, by which the mass accretion rate becomes as low as $10^{-3}
M_{\odot}$ s$^{-1}$, as shown in Fig.~\ref{fig2}(a) 
(Note that the mass accretion mainly comes from the region between the
rotational axis and the disk. See also Igumenshchev and Abramowicz
2000; Proga and Begelman 2003).

\placefigure{fig3}

We show in Fig.~\ref{fig4} profiles of mass fraction for
nuclear elements at the equatorial plane for model 0 at $t$ = 2.2 s. 
Dot-dashed, dotted, short-dashed, long-dashed, and solid lines represent mass
fraction of n, $\rm p$, $\rm ^4 He$, $\rm ^{16}O$, and $\rm ^{56}Ni$,
respectively. From this figure, we can easily see that oxygen is 
photo-dissociated into helium at $2 \times 10^8$cm, while helium is 
photo-dissociated into nucleons at $3 \times 10^7$cm. It is also noted, some
$\rm ^{56}Ni$ is seen at $ r \le 3 \times 10^7$ cm (solid line), which
may explain the luminosity of a hypernova as long as it is
ejected. This point is discussed in section~\ref{nucl-dis}.
The discontinuity of temperature at $3 \times 10^7$ cm in
Fig.~\ref{fig3} will come from the cooling effect due to the
photo-disintegration of helium into nucleons.

\placefigure{fig4}

From now on we show the results on neutrino processes. In
Fig.~\ref{fig5}, we show contour of neutrino cooling rate with
velocity fields for model 0 at $t$ = 2.2 s. The color represents the
emissivity of neutrinos (erg cm$^{-3}$ s$^{-1}$) in logarithmic scale
($10^{10}-10^{34}$). 
This emissivity of neutrinos is almost explained by
pair-captures on free nucleons, as shown in
Fig.~\ref{fig6}. 
We can easily see that emissivity of neutrinos is
high at the region where the accretion disk is formed.

\placefigure{fig5}

We show the results on neutrino cooling for every neutrino process. 
We show in Fig.~\ref{fig6}(a) cumulative energy ($\times 10^{50}$ erg) of
emitted neutrinos for each process as a function of time for model 0.
Dot-dashed, long-dashed, solid, short-dashed, and dotted lines
represent plasmon decay, electron-positron
pair annihilation, positron capture, electron capture,
and summation of all processes. It is clearly seen that
almost all emitted energy comes from pair captures on free nucleons. 
Also, the emitted energy amounts to of the order
of $10^{52}$ erg (strictly speaking, $3.44 \times 10^{52}$erg), which is
comparable to the accreted kinetic energy
and much higher than the
accreted thermal energy (see Fig.~\ref{fig2}(a)). Thus we
consider that almost all thermal energy, which was comparable to the
kinetic energy in the accretion disk, is extracted by neutrino emission.
Note that time evolution of electron fraction is 
mainly determined by 
positron and electron capture processes in the accretion disk.

\placefigure{fig6}

In Fig.~\ref{fig7}, we show contour of neutrino heating rate with
velocity fields for model 0 at $t$ = 2.2 s. The color represents the
energy deposition rate (erg cm$^{-3}$ s$^{-1}$) in logarithmic scale
($10^{10}-10^{34}$). Left panel shows the energy deposition rate due
to $\nu_e$ and $\bar{\nu}_e$ captures on free nucleons, while right
panel shows the energy deposition rate due to $\nu$ and $\bar{\nu}$
pair annihilation. In the pair annihilation, contributions from
three flavors are taken into account. It is, of course, that contour
of energy deposition rate due to $\nu_e$ and $\bar{\nu}_e$ captures
traces the number density of free nucleons, so the energy deposition
rate is high around the equatorial plane where the accretion disk is
formed. On the other hand, energy deposition rate due to $\nu$ and $\bar{\nu}$
pair annihilation occurs everywhere, including the region around the
polar axis. This feature will be good to launch a jet along the
polar axis, as pointed by MacFadyen and Woosley (1999). However, this
heating effect is too low to launch a jet in this study (see
Fig.~\ref{fig1}).

\placefigure{fig7}

In Fig.~\ref{fig8}(a), we show neutrino luminosity (solid line), energy
deposition rate due to $\nu$ and $\bar{\nu}$ pair annihilation
(dotted line), and energy deposition rate due to $\nu_e$ and
$\bar{\nu}_e$ captures on free nucleons (dashed line) as a function of time for
model 0. It is clearly seen that energy deposition rate is much
smaller than the neutrino luminosity, which supports our assumption
that the system is almost optically thin to neutrinos. Also, we can
see that $\nu_e$ and $\bar{\nu}_e$ captures on free nucleons
dominates $\nu$ and $\bar{\nu}$ pair annihilation process as the heating
process. It is also noted that neutrino luminosity and energy
deposition rate decreases along with time, which reflects
that the mass accretion rate also decreases along with time
(Fig.~\ref{fig2}(a)).

\placefigure{fig8}

Finally, we show in Fig.~\ref{fig9}(a) the integrated deposited energy
($\times 10^{50}$ erg ) due to $\nu$ and $\bar{\nu}$ pair
annihilation (solid line) and $\nu_e$ and $\bar{\nu}_e$ captures on
free nucleons (dashed line) as a function of time. It is confirmed 
that $\nu_e$ and $\bar{\nu}_e$ captures on free nucleons
dominate $\nu$ and $\bar{\nu}$ pair annihilation process
as the heating process. 
However, as shown in Figs.~\ref{fig5} and~\ref{fig7}, 
$\nu_e$ and $\bar{\nu}_e$ captures on free nucleons occur mainly
in the accretion disk, where neutrino cooling effect dominates neutrino
heating effect. Thus $\nu_e$ and $\bar{\nu}_e$ captures on free nucleons
is not considered to work for launching a jet. As for the
$\nu$ and $\bar{\nu}$ pair annihilation process, although this
process deposit energy everywhere, including the region around the
polar axis, the deposited energy amounts to only of the order of $10^{49}$
erg. This is $10^{-3}$ times smaller than the explosion energy of a
hypernova. 
In fact, as shown in Fig.~\ref{fig1}, the jet is not launched
in model 0. Thus we conclude that the efficiency of neutrino heating
is too low to launch a jet in this study.

\placefigure{fig9}

\subsection{Dynamics with Magnetic Fields}\label{withmag}

In this subsection, we explain the dynamics of model 9 as an example
of a collapsar with magnetic fields. Dependence of dynamics on the
initial amplitude of magnetic fields is shown in
section~\ref{dependence}. 

First, the accreted energy and mass accretion rate as a function of
time are shown in Fig.~\ref{fig2}(b). The meaning of each line is
same with Fig.~\ref{fig2}(a), although dot-dashed line, which is not
in Fig.~\ref{fig2} (a), represents accreted electro-magnetic energy
($\times 10^{50}$ erg) into the black hole as a function of time.
The reason why the final time of the simulation for model 9 is 2.23 s
is that the amplitude of the magnetic field (in particular,
$B_{\phi}$) becomes so high at the inner most region
that the Alfv\'{e}n crossing time 
at the region make the time step extremely small.
Also, it seems that the mass accretion
rate does not decrease so much in model 9. Rather, it seems to keep
$\sim 0.05M_{\odot}$ s$^{-1}$. We guess this is because
magnetic fields play a role to transfer the angular momentum from
inner region to outer region, which makes matter fall into the black
hole more efficiently.
This point is also discussed with Figs.~\ref{fig8}---\ref{fig12}.

Next, we show in Fig.~\ref{fig10} profiles of physical quanta of the
accretion disk around the equatorial plane for model 9 at $t$ = 2.2
s. When we
compare these profiles with the ones in Fig.~\ref{fig3}, we can see
that the radial velocity is higher in model 9 at small radius. 
This may reflect that the mass accretion rate is higher in model 9 than
model 0 
(as stated in section 3.1, the
mass accretion mainly comes from the region between the
rotational axis and the torus. So we have to note that this feature does not
explain the mass accretion rate directly. see also Fujimoto et
al. 2006).
As for the other profiles, there seems no significant
difference between the two models.

\placefigure{fig10}

\begin{figure}
\epsscale{1.0}
\plotone{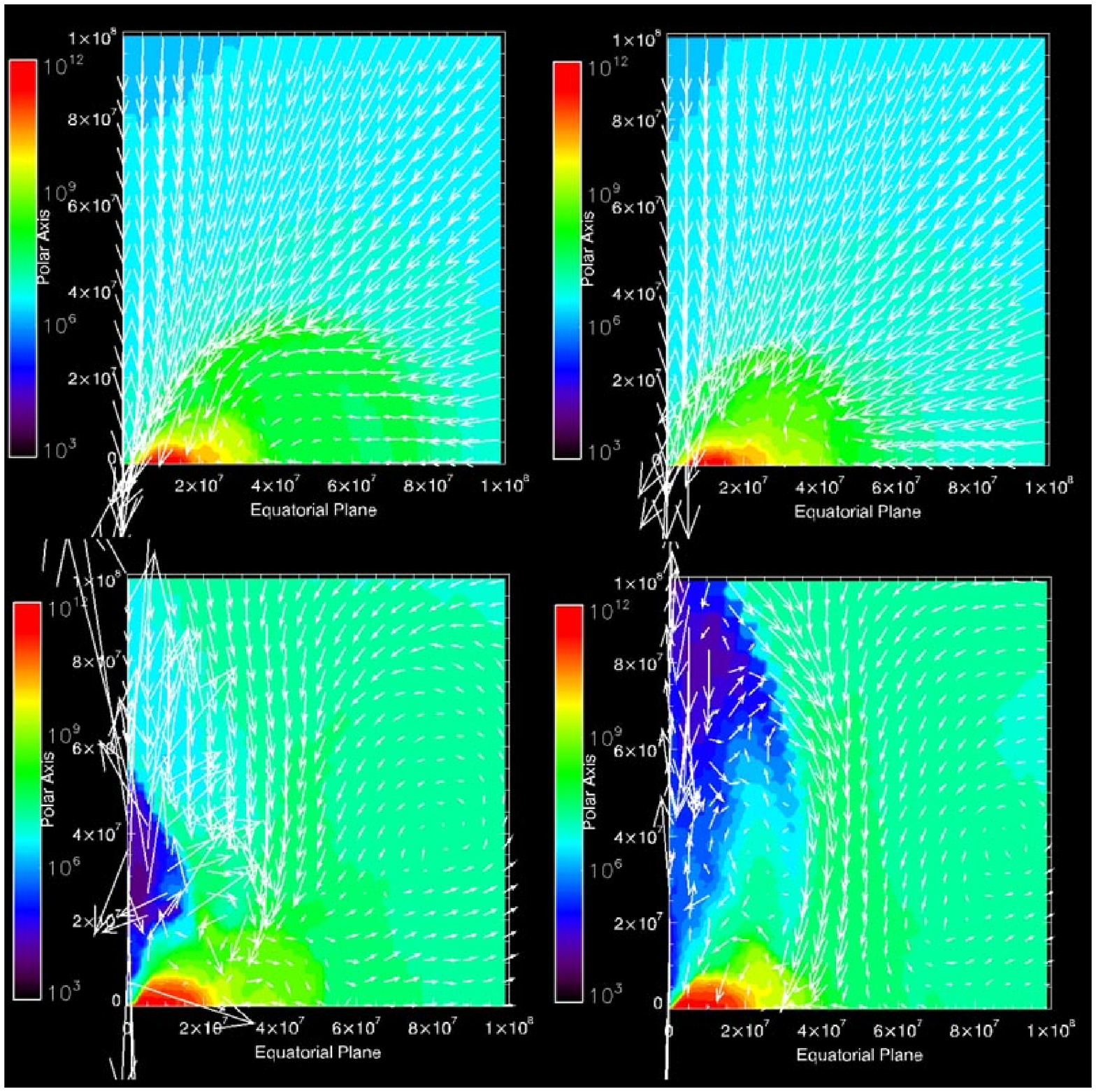}
\caption{Top panels: density contour with velocity fields for model 0
at $t$ = 2.1 s (left panel) and $t$ = 2.2 s (right panel). Bottom
panels: same with upper panels, but for model 9.
The color represents the density (g cm$^{-3}$) in logarithmic scale
($10^{3}-10^{12}$). 
In this figure, the central region of the progenitor ($r \le 10^8$cm)
is shown. Vertical axis and horizontal axis represent rotation axis
and equatorial plane, respectively.
 \label{fig1}}
\end{figure}

\begin{figure}
\epsscale{1.0}
\plotone{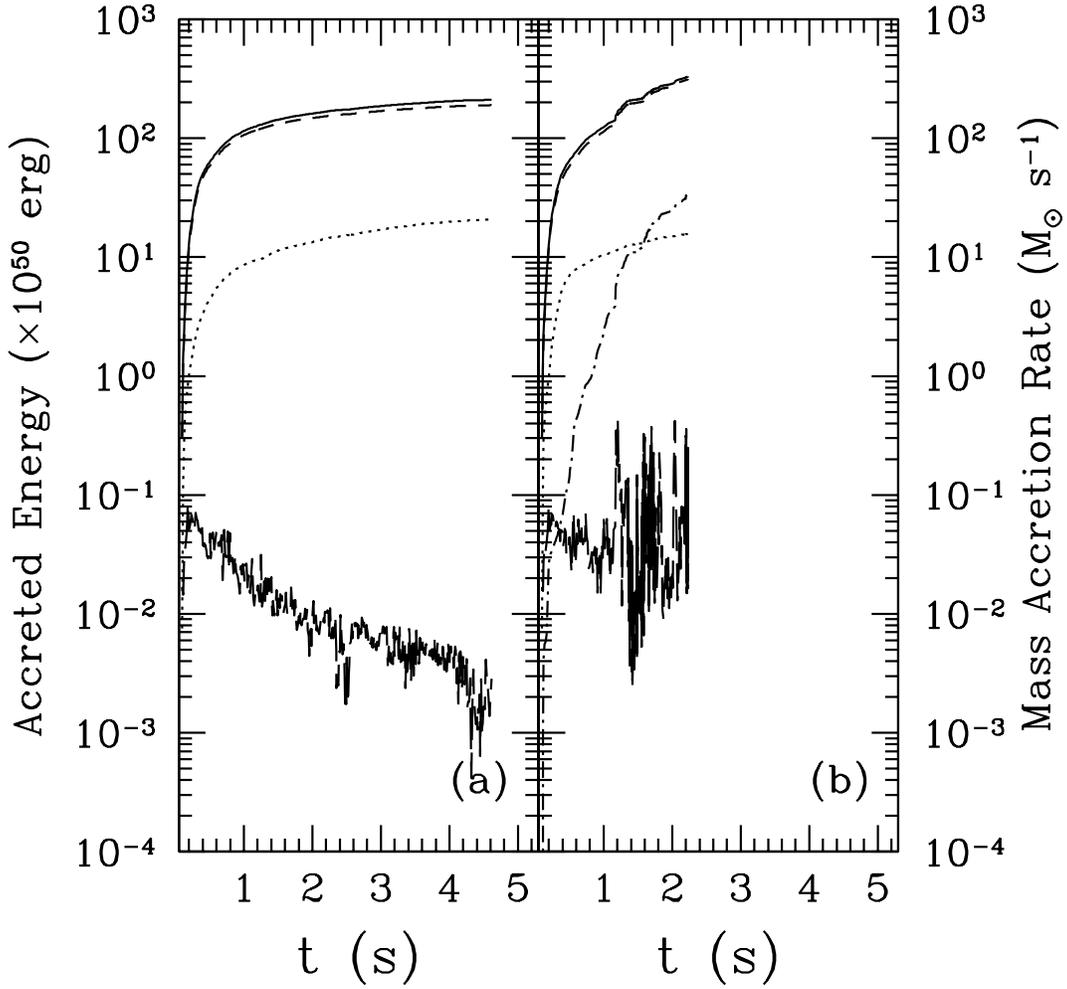}
\caption{Solid, short-dashed, dotted, dot-dashed lines represent
accreted total energy, kinetic energy, thermal energy, and
electro-magnetic energy ($\times 10^{50}$ erg) into the black hole as
a function of time, respectively. Long-dashed lines represent the
mass accretion rate ($M_{\odot}$ s$^{-1}$) as a function of time.
Left panel shows the case for model 0, while right panel represents
the case for model 9. Note that in the case of model 0, the
electro-magnetic energy is set to be zero.
 \label{fig2}}
\end{figure}

\begin{figure}
\epsscale{1.0}
\plotone{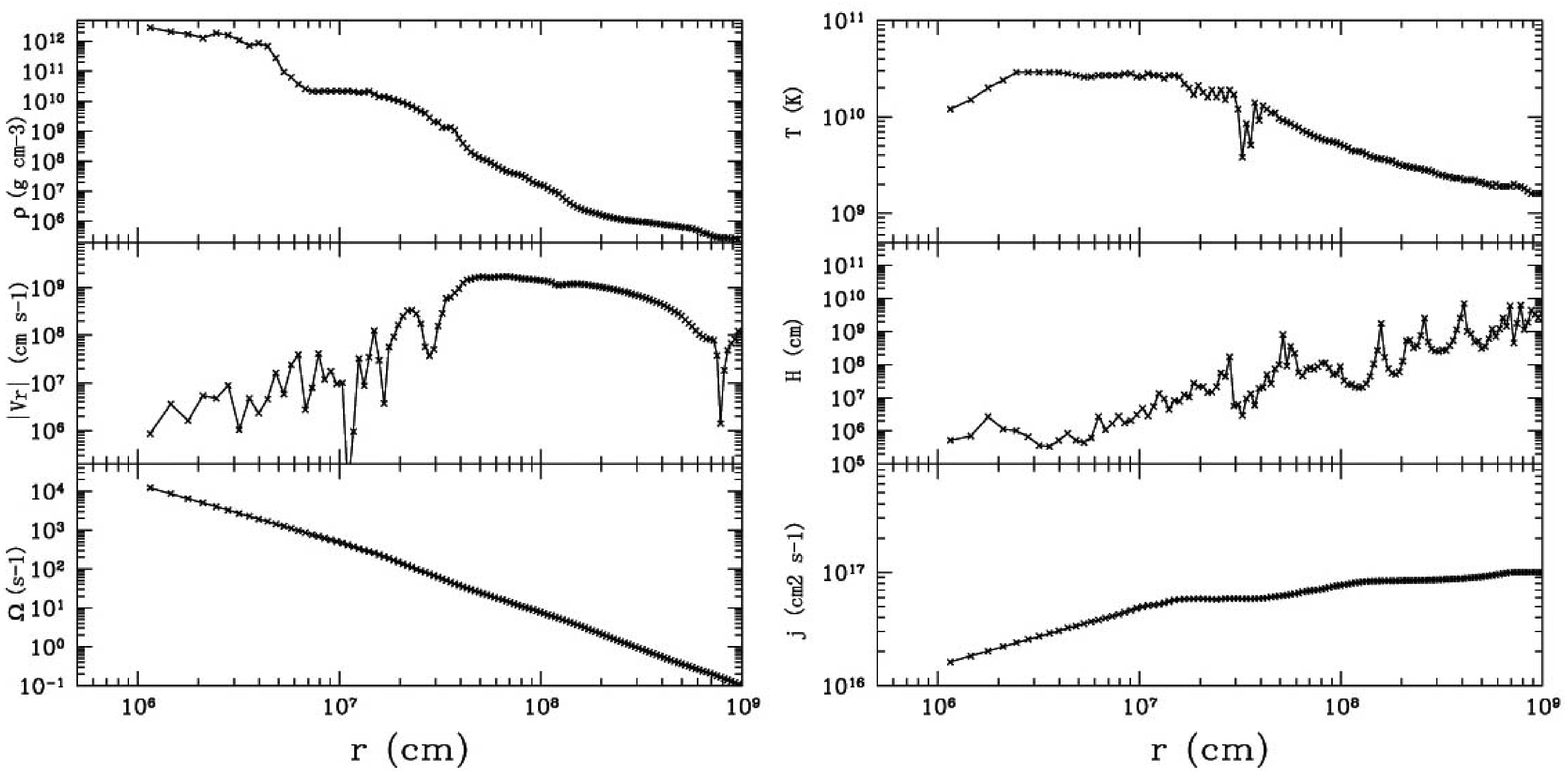}
\caption{Profiles of density, absolute value of radial velocity, angular
frequency, temperature, density scale height, and specific angular momentum
at the equatorial plane for model 0 at $t$ = 2.2 s. 
\label{fig3}}
\end{figure}

\begin{figure}
\epsscale{1.0}
\plotone{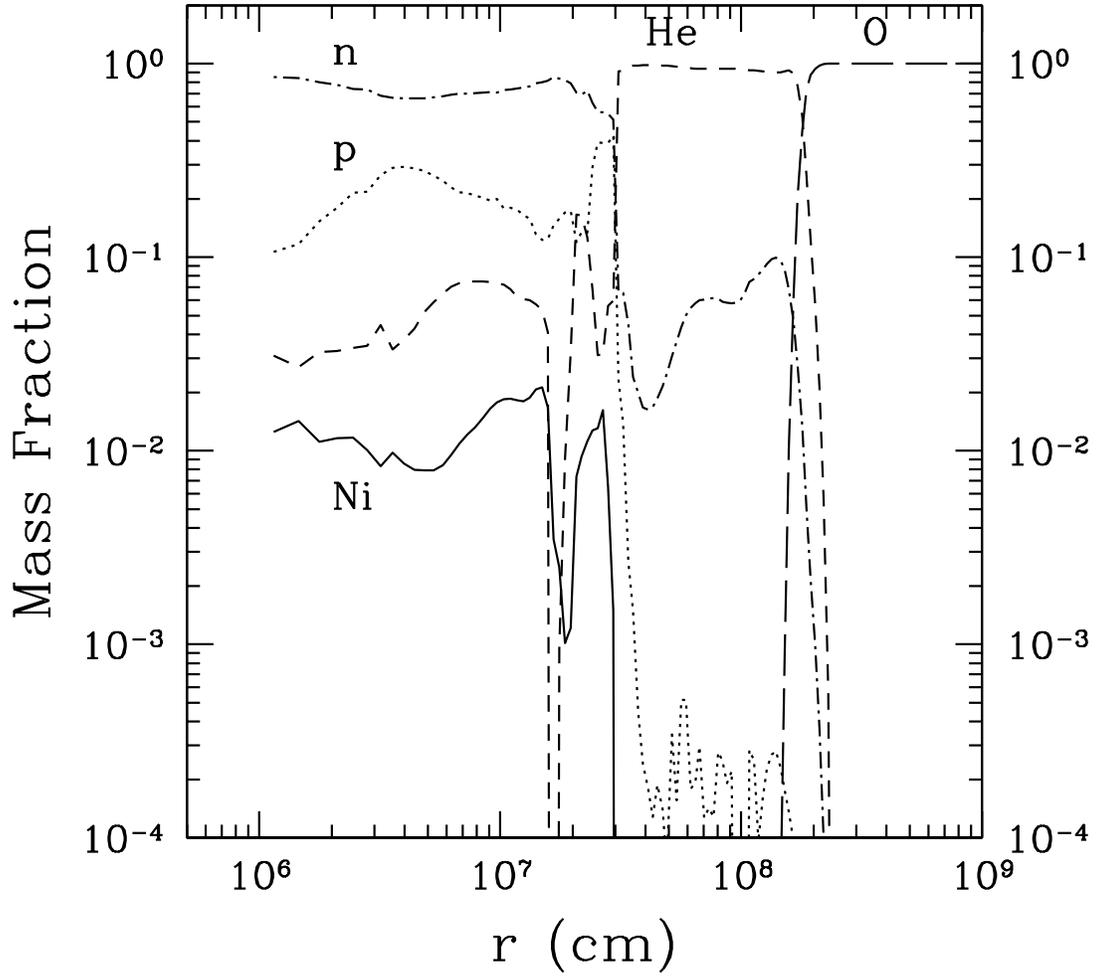}
\caption{Profiles of mass fraction for nuclear elements at the
equatorial plane for model 0 at $t$ = 2.2 s. Dot-dashed, dotted,
short-dashed, long-dashed, and solid lines represent mass
fraction of n, $\rm p$, $\rm ^4 He$, $\rm ^{16}O$, and $\rm ^{56}Ni$,
respectively.
\label{fig4}}
\end{figure}

\begin{figure}
\epsscale{1.0}
\plotone{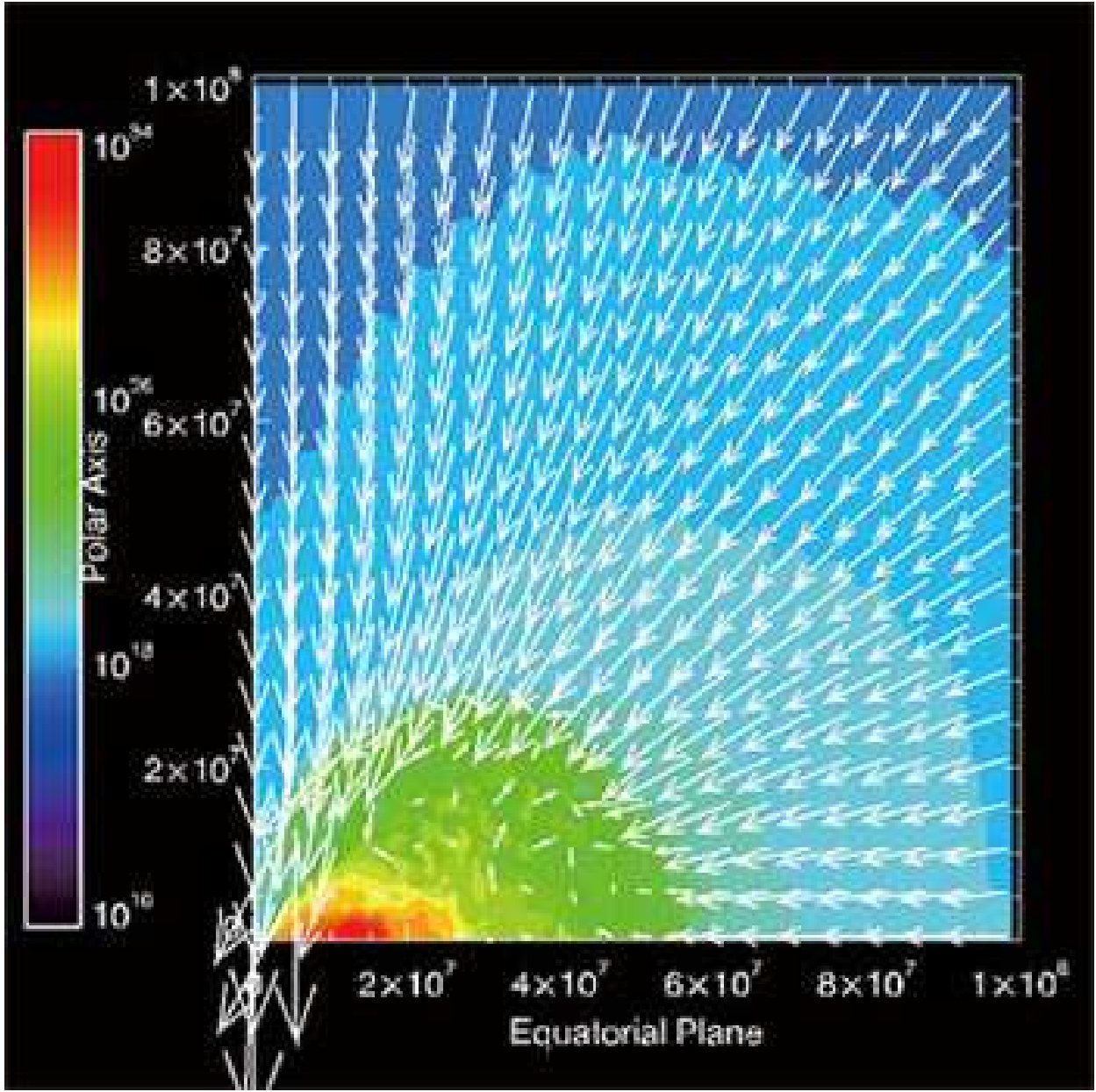}
\caption{Contour of neutrino cooling rate with velocity fields for model 0
at $t$ = 2.2 s. The color represents the emissivity of neutrinos
(erg cm$^{-3}$ s$^{-1}$) in logarithmic scale ($10^{10}-10^{34}$).
\label{fig5}}
\end{figure}

In Fig.~\ref{fig11}, we show profiles of amplitude of magnetic fields
as a function of radius on the equatorial plane
for model 9 at $t$ = 2.2 s. Dotted, solid, and dashed lines
represent the amplitude of $B_r$, $B_\theta$, and $B_\phi$,
respectively. It is clearly seen that $B_\phi$ dominates within
$r=10^8$cm. Thus we can conclude that magnetic pressure from $B_\phi$
drives the jet along the rotation axis (see Fig.~\ref{fig1}). This
point is also discussed with Figs.~\ref{fig13}---\ref{fig15}. Also,
these magnetic fields may play a role to transfer the angular
momentum. Since the viscosity parameter $\alpha$ can be estimated
as~\citep{balbus98,akiyama03} 
\begin{eqnarray}
\alpha \sim \frac{B_r B_\phi}{4 \pi P},
\label{Eq:alpha} 
\end{eqnarray}
we plot in Fig.~\ref{fig11} the estimated viscosity 
parameter ($\times 10^{-20}$)
(long-dashed line). 
This figure suggests that $\alpha$-viscosity
coming from the magnetic fields may play a role to transfer the
angular momentum at inner most region effectively. 
However, the angular momentum cannot be transfered to infinity
along the radial direction. 
This is confirmed by the Alfv\'{e}n
Mach number ($\times 10^{-10}$) in the radial 
direction ($\equiv v_r/v_{r,A}$; dot-dashed line
in Fig.~\ref{fig11}). 
At only the inner most region, the flow becomes marginally
sub-Alfv\'{e}nic where the viscous force due to magnetic stress can bring the
angular momentum outward. Thus we consider that the outflow (including the jet)
in the polar direction (see Fig.~\ref{fig1} bottom) should bring the angular 
momentum from the inner most region.

As additional information, we found that the velocity of the
slow magnetiosonic wave is almost same with the Alfv\'{e}n velocity.
On the other hand, we found that the fluid is subsonic against the
fast magnetiosonic wave in the simulated region.     

\placefigure{fig11}

We show in Fig.~\ref{fig12}
the profiles of specific angular momentum (cm$^2$ s$^{-1}$)
on the equatorial plane
for model 0 (dotted line) and model 9 (short-dashed line) at $t$ = 2.2 s. 
The profiles of angular momentum density (g cm$^{-1}$ s$^{-1}$)
for model 0 (long-dashed line) and model 9 (solid line)
are also shown in the figure. We confirmed that the specific
angular momentum is not so different between model 0 and model 9,
but it is found that the angular momentum density is lower
in model 9 compared with model 0 at the inner region.
This feature might reflect that the matter falls into the black hole
efficiently in model 9 at small radius (see Fig.~\ref{fig11}).
This picture seems to be consistent with the almost 
constant accretion rate in model 9 (Figs.~\ref{fig2}(b)).

\placefigure{fig12}


Cumulative energy ($\times 10^{50}$ erg) of emitted
neutrinos for each process as a function of time for model 9 is shown in
Fig.~\ref{fig6}(b), 
neutrino luminosity, energy deposition rate due
to $\nu$ and $\bar{\nu}$ pair annihilation, and energy
deposition rate due to $\nu_e$ and $\bar{\nu}_e$ captures on free
nucleons as a function of time for model 9 are shown in Fig.~\ref{fig8}(b), 
and integrated deposited energy due to
$\nu$ and $\bar{\nu}$ pair annihilation and
$\nu_e$ and $\bar{\nu}_e$ captures on free nucleons as
a function of time for model 9 are shown in Fig.~\ref{fig9}(b)
The meaning of each line in each figure is same with the one used for
model 0. 
We can derive a similar conclusion for the role of neutrino heating
effect in model 9 as in model 0. As for the $\nu_e$ and $\bar{\nu}_e$ 
captures on free nucleons, the cumulative deposited energy becomes
as high as $10^{52}$ ergs that is comparable to the explosion energy
of a hypernova. However, this heating process occurs mainly
in the accretion disk, where neutrino cooling effect dominates neutrino
heating effect. Thus $\nu_e$ and $\bar{\nu}_e$ captures on free nucleons
is not considered to work for launching a jet. As for the
$\nu$ and $\bar{\nu}$ pair annihilation process, 
the deposited energy amounts to no more than $10^{50}$ erg. 
Thus we conclude that the neutrino heating effect in model 9 
is too inefficient to launch a GRB jet and cause a hypernova.
It has to be noted that the energy deposition rate due to 
pair captures on free nucleon sometimes becomes larger than
the neutrino luminosity in Fig.~\ref{fig8}(b). This means that
the optically thin limit breaks down at that time. This point is
also discussed in section~\ref{neutrino-dis}.
We also found that the neutrino luminosity,
energy deposition rate, and integrated deposited energy seem to be
higher in model 9 than in model 0. We consider that this feature
comes from the high mass
accretion rate (high rate of release of gravitational energy) caused
by angular momentum transfer due to magnetic fields.

We show in Fig.~\ref{fig13} time evolution of energy of magnetic fields
($\times 10^{50}$ erg) in the whole calculated region
for the case of model 9. Dotted, solid, and
dashed lines represent energy in the form of $B_r$, $B_{\theta}$, and
$B_{\phi}$, respectively. We can see that $B_{\phi}$ grows rapidly and
dominates other components, although $B_{\phi}$ is much smaller than
other components at first (note that 1-D simulation of the spherical
collapse of the progenitor is done until $t=$0.1 s (see section 2.2), 
so $B_{\phi}$ is set to be 0 until $t=0.1$ s). 
$B_{\phi}$ is amplified by winding effect
and launches a jet along the polar axis (see Fig.~\ref{fig1}).
The total energy of $B_{\phi}$ at the final stage of the simulation in
this study is of the order of $10^{50}$ erg, which is much smaller
than the explosion energy of a hypernova. However, the total energy of  
$B_{\phi}$ is increasing almost monotonically. As mentioned above,
the final time of the simulation for model 9 is determined 
due to the reason that the Alfv\'{e}n crossing time becomes extremely small
at the inner most region.
When we can overcome this problem and simulate the model 9 further,
there is a possibility that the energy of $B_{\phi}$ becomes
comparable to the explosion energy of a hypernova. The evolution
of $B_{r}$ and $B_{\theta}$ is similar to the one by Magnetic Rotational
Instability (MRI), which is discussed in section~\ref{b-fields-dis}.

\placefigure{fig13}

We show in Fig.~\ref{fig14} 
contour of plasma beta (=$p_{\rm gas+radiation}/p_{\rm mag}$)
with magnetic fields ($B_{r}$ and $B_{\theta}$) for model 9
at $t$ = 2.2 s. In $p_{\rm gas+radiation}$, degenerated pressure
of electrons is included. The color represents the plasma beta
in logarithmic scale ($10^{-1}-10^{8}$). 
The minimum value of plasma beta in this region is 0.193.
We can see clearly that the beta value is low in the jet region,
from which we can understand the jet is launched by the magnetic
fields (in particular, $B_{\phi}$). 

\placefigure{fig14}

Finally, we have performed these simulations presented
in sections~\ref{withoutmag} and~\ref{withmag} without calculating
self-gravity, and found that the dynamics of the collapsar is 
not changed so much.

\subsection{Toward Discussions}\label{toward}

Before we discuss the results mentioned above, we present two more
results in this section. One is the dependence of dynamics on initial
amplitude of magnetic fields, the other is the explosive
nucleosynthesis.  

\subsubsection{Dependence on Initial Amplitude of Magnetic
Fields and on Resolution of Grids}\label{dependence} 

Here we show the dependence of dynamics on initial
amplitude of magnetic fields. In Fig.~\ref{fig15}, evolutions of 
total energy
of magnetic fields ($\times 10^{50}$ erg) for model 8 (top-left
panel), model 10 (top-right panel), model 11 (bottom-left panel), and
model 12 (bottom right panel) are shown. The final time of the simulation
is determined, like in model 9, by the reason that the Alfv\'{e}n crossing
time becomes extremely small at the inner most region. In all models,
the energy of $B_\phi$ amounts to $10^{49} - 10^{50}$ erg, which
means that the amplitude of $B_\phi$ becomes as strong as
$10^{15}$G. This can be understood by simple calculations: the typical
radius of the inner most region is of the order of $10^6$ cm, so the
volume of the region times the energy density of magnetic fields becomes
$E_{B} = 8.3 \times 10^{49} (B/2\times 10^{15} \rm G)^2 
(r/5\times 10^6 \rm cm)^3 $ erg.

\placefigure{fig15}

The total energy of $B_r$ and $B_\theta$ components does not change so much with
time. In particular, in models 10---12, their total energy hardly
changes. When we compare $B_r$ with $B_\theta$, $B_r$ seems to be more
unstable than $B_\theta$ (see also Fig.~\ref{fig13}). This is similar to
the results of the local simulations of MRI,
which is discussed in section~\ref{b-fields-dis}.

We summarize the features of the launched jet at the final stage of
simulations for every model in Table~\ref{tab2}. We define the jet as
follows: (i) the region within some opening angle ($\theta_j$)
(ii) the region where total energy (i.e. summation of kinetic,
thermal, electro-magnetic, and gravitational energies) is positive
at the final stage of the simulations. (iii) the amplitude of the
velocity is larger than $5\times 10^9$ cm s$^{-1}$.
In Table~\ref{fig2}, total
energy, mass, terminal bulk Lorentz factor, and ratio of the magnetic
energy relative to total energy of the jet is shown assuming that
$\theta_J$ = 5$^\circ$, 10$^\circ$, and 15$^\circ$ for each model.
Terminal bulk Lorentz factor is estimated by assuming that total
energy goes into kinetic energy during expansion. 

We can see that terminal bulk Lorentz factor in every model is much
smaller than the required value ($\sim 300$) for GRB jets. 
Also, the energy of the jet in every model is much smaller than 
the typical energy of a GRB.
Thus we conclude that the jets seen in this study will not be GRB jets.

\placetable{tab2}

Also, we show in left panel of Fig.~\ref{fig16} the
evolution of total energy of magnetic fields for the case of model 9
(same with Fig.~\ref{fig13}), but for 150($r$)$\times$5($\theta$) grid
points (left panel) and 150($r$)$\times$20($\theta$) grid points (right
panel) to show the dependence of results on the grid resolution. 
It is found that the growth rate of $B_{r}$ and $B_{\theta}$ components
depends on the grid resolution. 
On the other hand the growth rate of $B_{\phi}$ does not
depend on the grid resolution (that is, the energy in $B_{\phi}$
fields becomes $10^{49}-10^{50}$ ergs in (1-2) sec). 
The saturation level of $B_{r}$ and $B_{\theta}$
does not depend on the grid resolution so much.
This is discussed in section~\ref{b-fields-dis}.
We have also done a simulation of model 9 with finer
resolution (300($r$)$\times$60($\theta$)), although neutrino
anti-neutrino pair annihilation is not included to save CPU time.
The simulation region is set to be (10$^6$ cm $\le$ $r$ $\le$ $10^9$
cm, 0 $\le$ $\theta$ $\le$ 90). The minimum radial grid is set to be
3$\times 10^5$ cm. We have found that the dynamics of collapsars is
hardly changed. We have found that an accretion disk is formed
around the black hole and a jet is launched at $t=1.98$ sec. Moreover,
we have found that the evolution of magnetic fields is hardly changed,
which is shown in right panel of Fig.~\ref{fig16}. This result means that 
a jet is driven by magnetic fields and the standard resolution of
our study (150($r$)$\times$30($\theta$)) is not so bad.

\placefigure{fig16}

\subsubsection{Nucleosynthesis}\label{nucleosynthesis}

In this section, we present results on explosive nucleosynthesis.
We show in Table~\ref{tab3} the mass of $\rm ^{56}Ni$ in the regions 
where total energy is positive ($M_{\rm Ni}^{\rm esc}$ ($M_{\odot}$))
and 
the total mass of $\rm ^{56}Ni$ in the whole simulated region
($M_{\rm Ni}^{\rm tot}$ ($M_{\odot}$)) for each model. These amounts are
estimated at the final stage of the simulations. 
The ejected mass of $\rm ^{56}Ni$, $M_{\rm Ni}^{\rm esc}$, is too
little to explain the luminosity of a hypernova, although 
considerable amount of $\rm ^{56}Ni$ is synthesized
in the accretion disk ($M_{\rm Ni}^{\rm tot}$ ($M_{\odot}$)).
We found that the ejected mass is mainly composed of $n$, $p$, 
and $\rm He$. This is supported by Fig.~\ref{fig17} where entropy
per baryon at $t=2.2$ s for model 9 is shown in units of $k_b$. 
In the jet region, the entropy
per baryon is remarkably high, so it is natural that light elements
dominate in the jet region.

\placetable{tab3}
\placefigure{fig17}

Finally, we show in Fig.~\ref{fig18} contour of electron fraction
($\rm Y_e$) with
velocity fields for model 0 at $t$ = 2.2 s (left panel) and
model 9 at $t$ = 2.2 s (right panel). The color represents the
electron fraction on a linear scale (0.1-0.540).
We can easily see that $\rm Y_e$ becomes low in the accretion disk.
This is because electrons are degenerated and 
electron capture dominates positron capture at this region. 
Also, we can see that mass element with low $Y_e$ is ejected in model
9. Also, mass element with high $Y_e$ (highest value is 0.522)
is also ejected in model 9 from the inside of the low $Y_e$ jet
along the polar axis near the black hole.

Since entropy per baryon is
very high in the jet region (Fig.~\ref{fig17}),
these mass elements may cause
$r-$process and/or $r/p-$process nucleosynthesis. 
Here we have to comment on the electron fraction at the high density
region. The value of the electron fraction is solved of the order of
0.1 in the accretion disk. 
This value is obtained assuming that the chemical potential of
electron-type neutrinos is zero.
If careful treatment of neutrino transfer is done,
the chemical potential of electron-type neutrinos may prohibit
the electron fraction from being as low as 0.1. This point is
discussed in sections~\ref{neutrino-dis} and~\ref{nucl-dis}.

\placefigure{fig18}

\section{DISCUSSIONS}\label{discussions}

In this section, we discuss our numerical results and prospect for
future works. We discuss neutrino physics, effects of magnetic fields,
nucleosynthesis, general relativistic effects, initial conditions, and
prospect for improvements of our numerical code.

\subsection{Neutrino Physics}\label{neutrino-dis} 

\begin{figure}
\epsscale{1.0}
\plotone{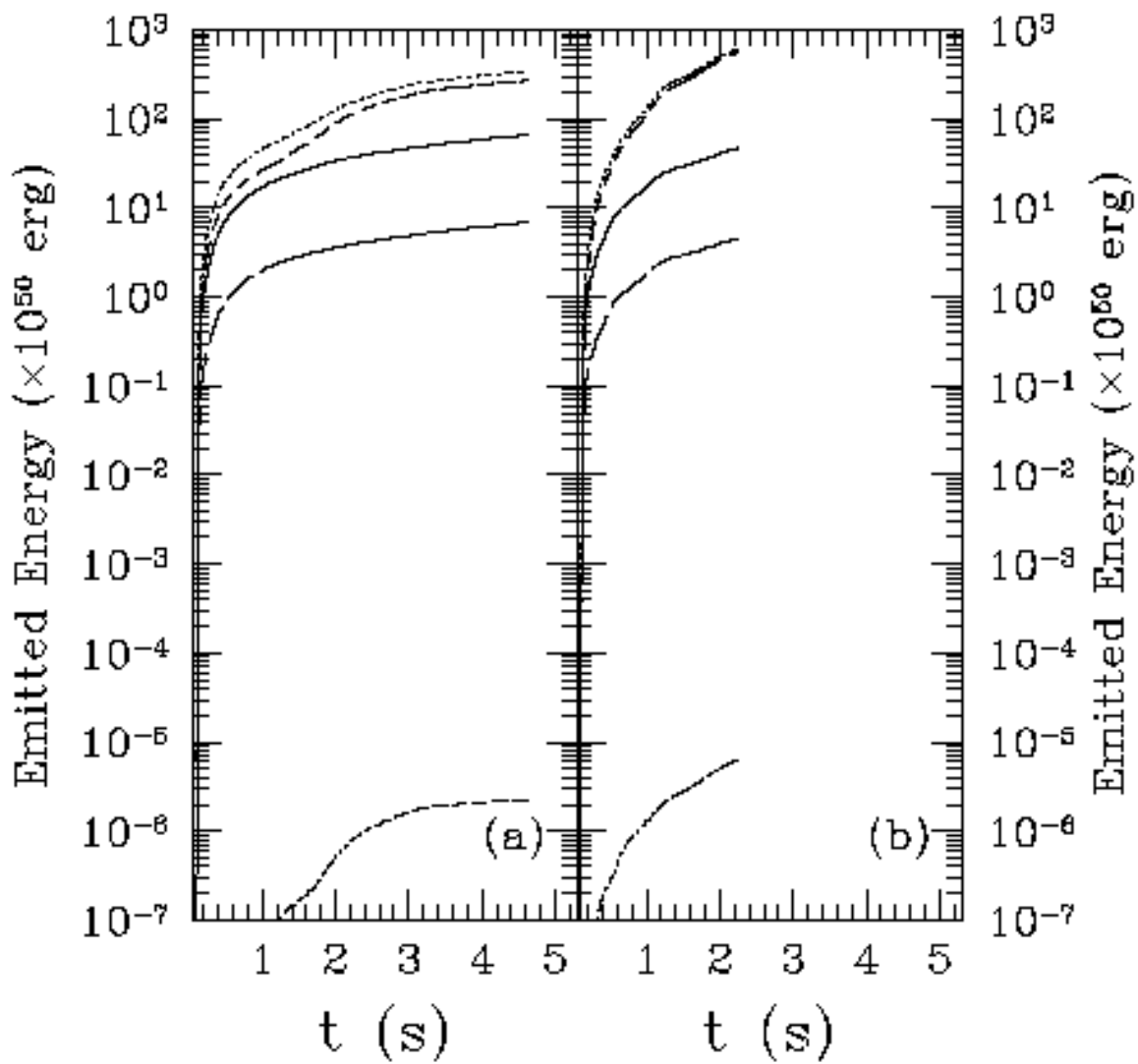}
\caption{Cumulative energy ($\times 10^{50}$ erg) of emitted
neutrinos for each process as a function of time. 
Dot-dashed, long-dashed, solid, short-dashed, and dotted lines represent
plasmon decay, electron-positron pair annihilation, positron capture,
electron capture, and summation of all processes. Left panel shows
the case for model 0, while right panel shows model 9.
 \label{fig6}}
\end{figure}

\begin{figure}
\epsscale{1.0}
\plotone{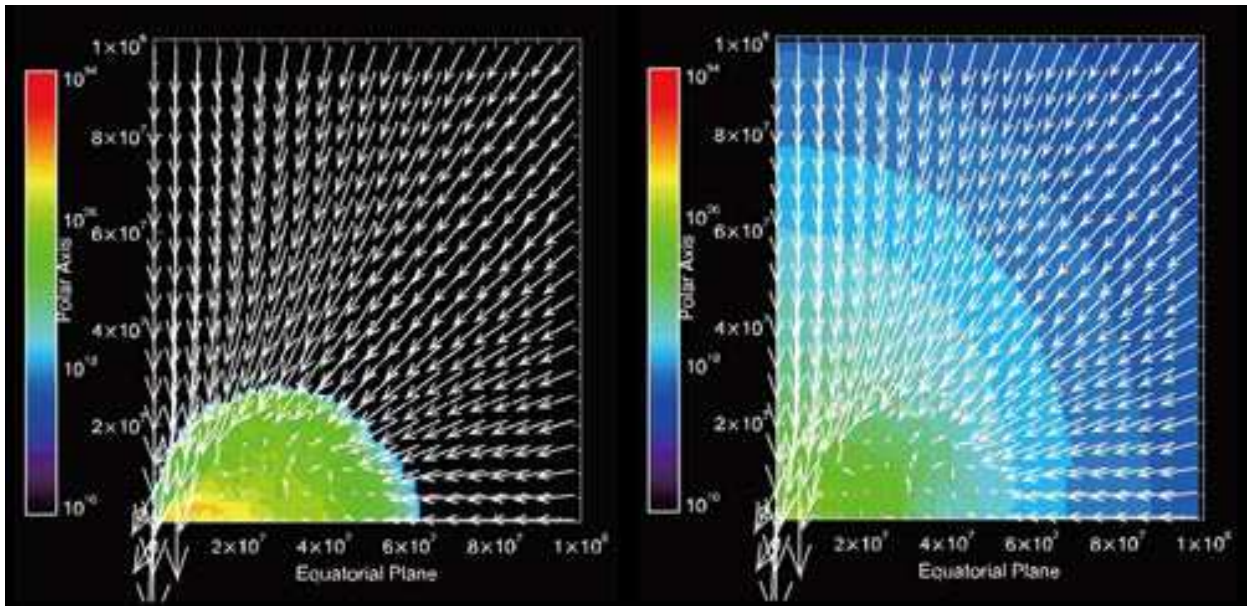}
\caption{Contour of neutrino heating rate with velocity fields for model 0
at $t$ = 2.2 s. The color represents the energy deposition rate
(erg cm$^{-3}$ s$^{-1}$) in logarithmic scale ($10^{10}-10^{34}$). 
Left panel shows the
energy deposition rate due to $\nu_e$ and $\bar{\nu}_e$ captures on
free nucleons, while right panel shows the energy deposition rate due
to $\nu$ and $\bar{\nu}$ pair annihilations (three flavors are taken
into account).
 \label{fig7}}
\end{figure}

\begin{figure}
\epsscale{1.0}
\plotone{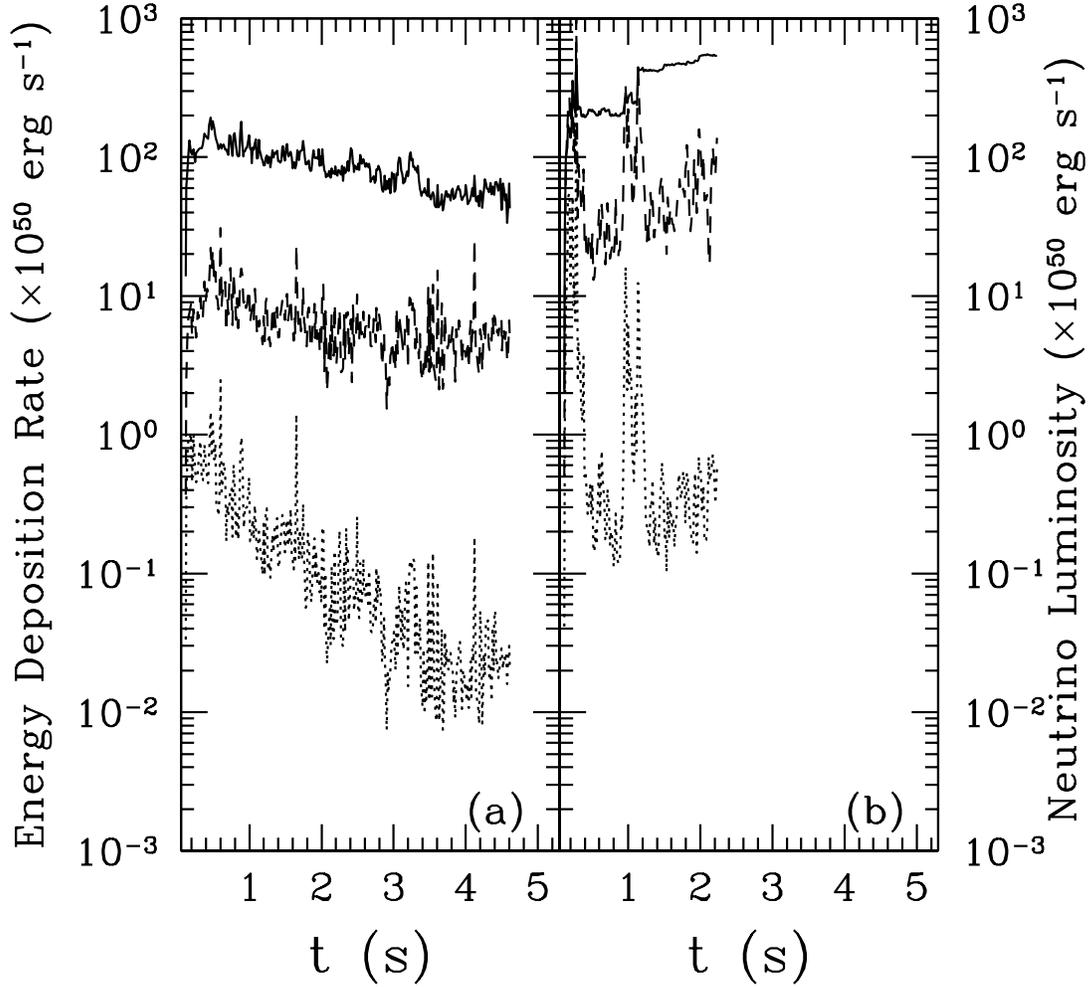}
\caption{Neutrino luminosity (solid lines), energy deposition rate due
to $\nu$ and $\bar{\nu}$ pair annihilations (dotted lines), and energy
deposition rate due to $\nu_e$ and $\bar{\nu}_e$ captures on free
nucleons (dashed line) as a function of time. The unit is $10^{50}$ erg
s$^{-1}$. Left panel shows the case for model 0, while right panel
shows the case for model 9.  
\label{fig8}}
\end{figure}

\begin{figure}
\epsscale{1.0}
\plotone{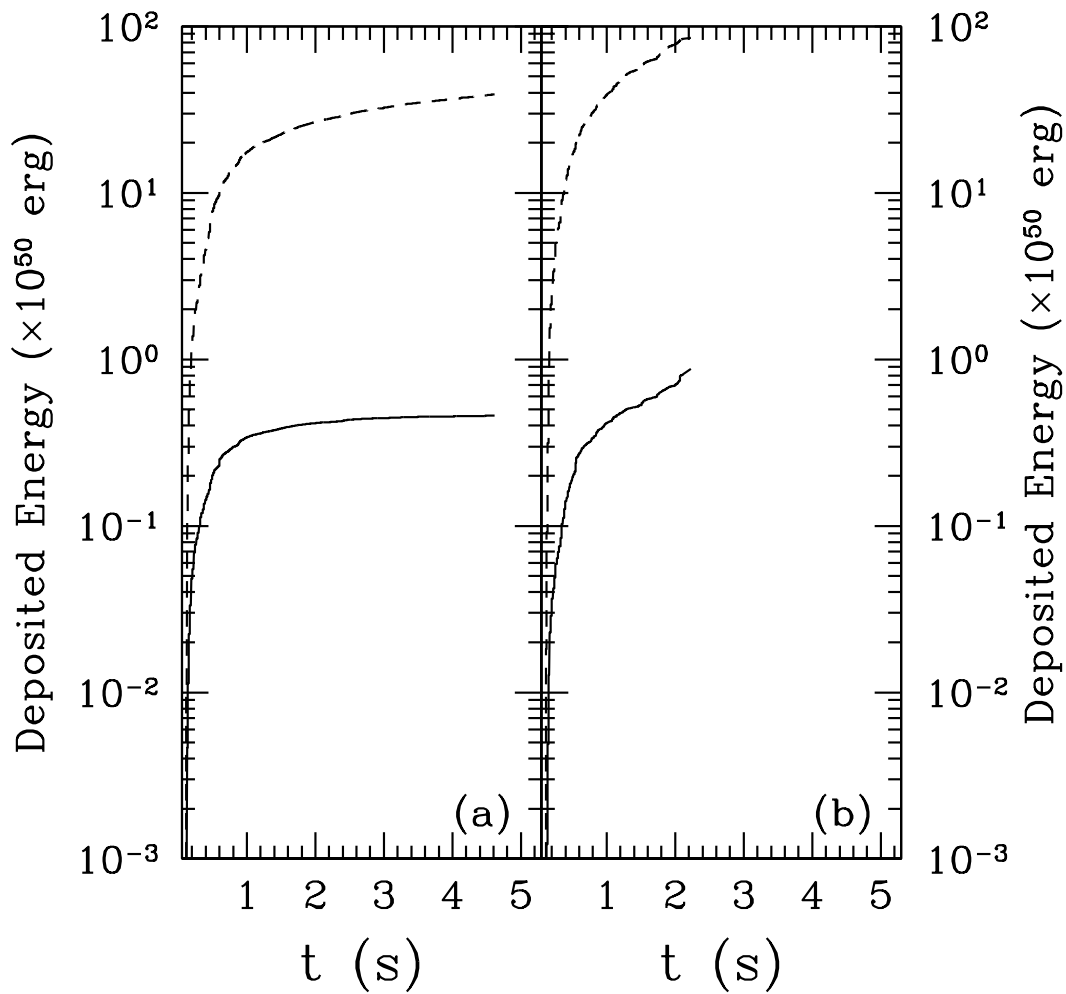}
\caption{Integrated deposited energy ($\times 10^{50}$ erg ) due to
$\nu$ and $\bar{\nu}$ pair annihilations (solid lines) and
$\nu_e$ and $\bar{\nu}_e$ captures on free nucleons (dashed lines) as
a function of time. Left panel shows the case for model 0, while right panel
shows the case for model 9. 
 \label{fig9}}
\end{figure}

\begin{figure}
\epsscale{1.0}
\plotone{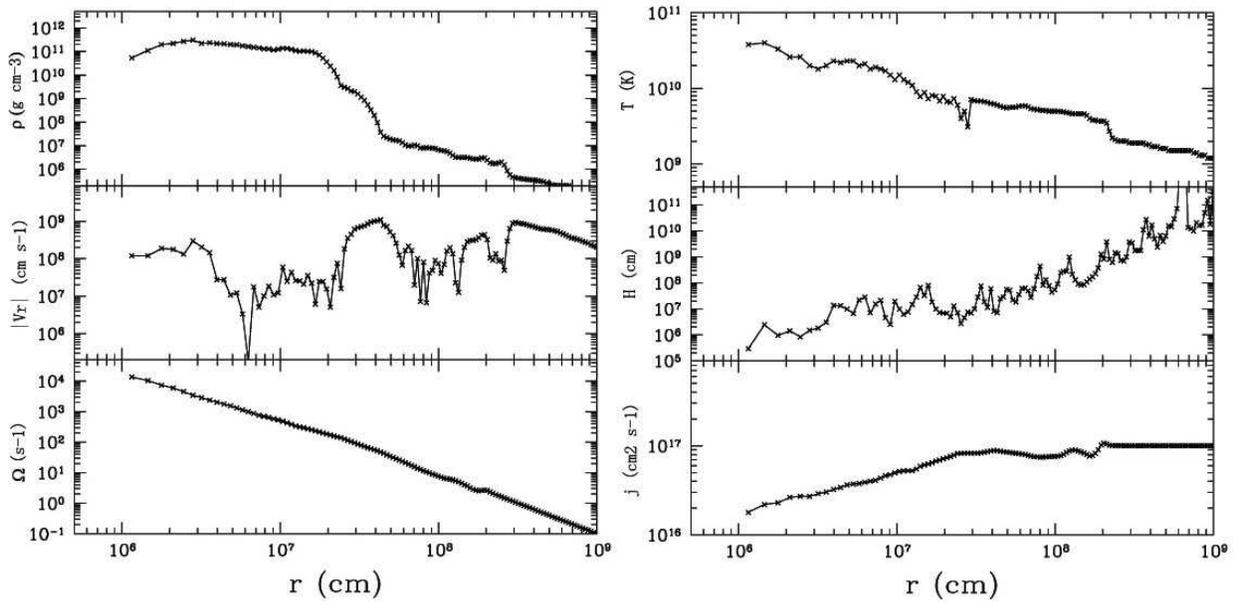}
\caption{
Same with Fig.3, but for model 9. Note that the radial velocity
profile is not similar to model 0 at small radius. 
 \label{fig10}}
\end{figure}

\begin{figure}
\epsscale{1.0}
\plotone{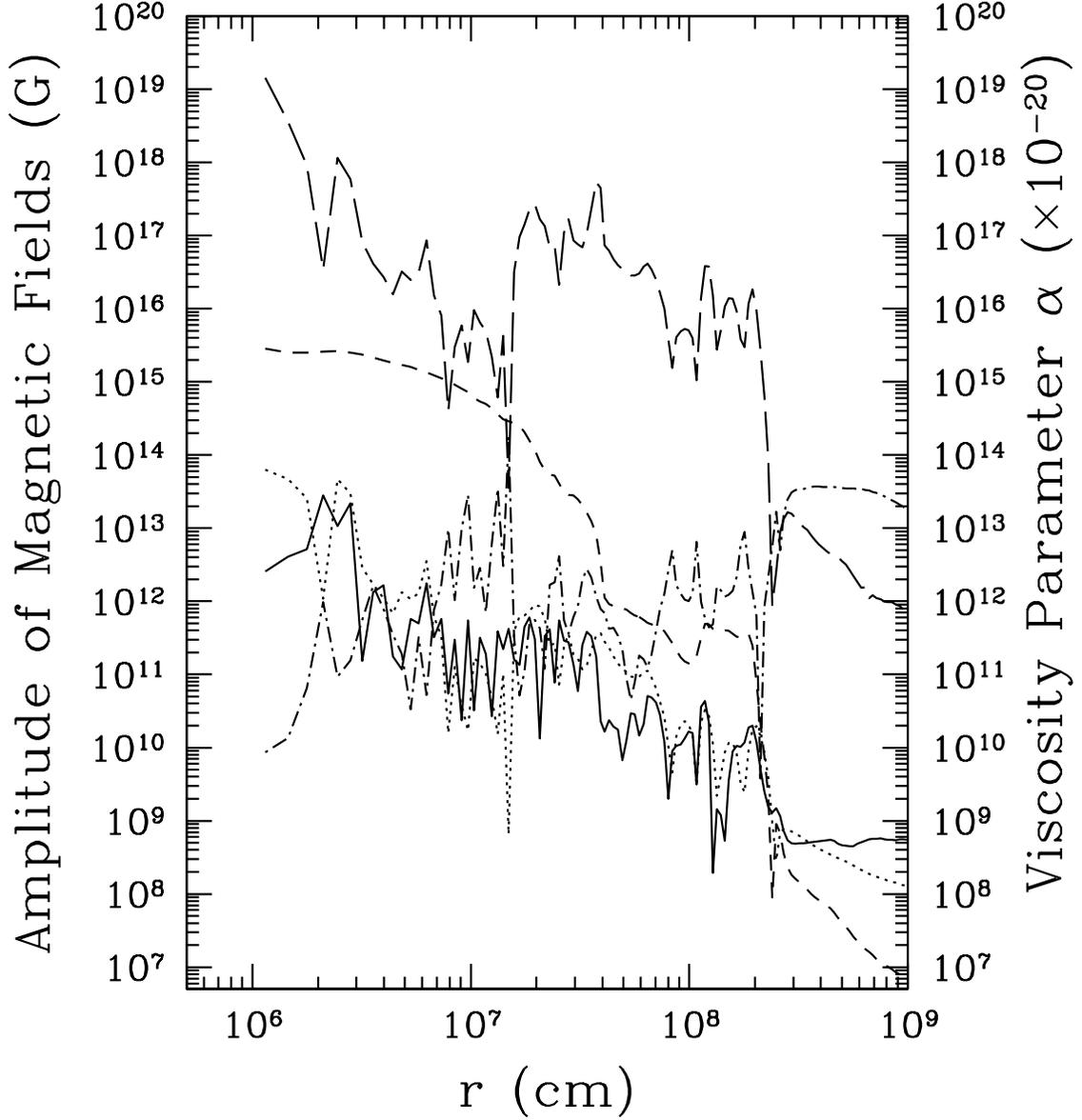}
\caption{Profile of amplitude of magnetic fields and estimated
viscosity parameter $\alpha$ ($\times 10^{-20}$)
on the equatorial plane for model 9 at $t$ = 2.2
s. Dotted, solid, and short-dashed lines represent the amplitude of
$B_r$, $B_\theta$, and $B_\phi$, while long-dashed line represent the
estimated viscosity parameter. Dot-dashed line represents the Alfv\'{e}n
Mach number ($\times 10^{-10}$) in the radial direction. 
 \label{fig11}}
\end{figure}

\begin{figure}
\epsscale{1.0}
\plotone{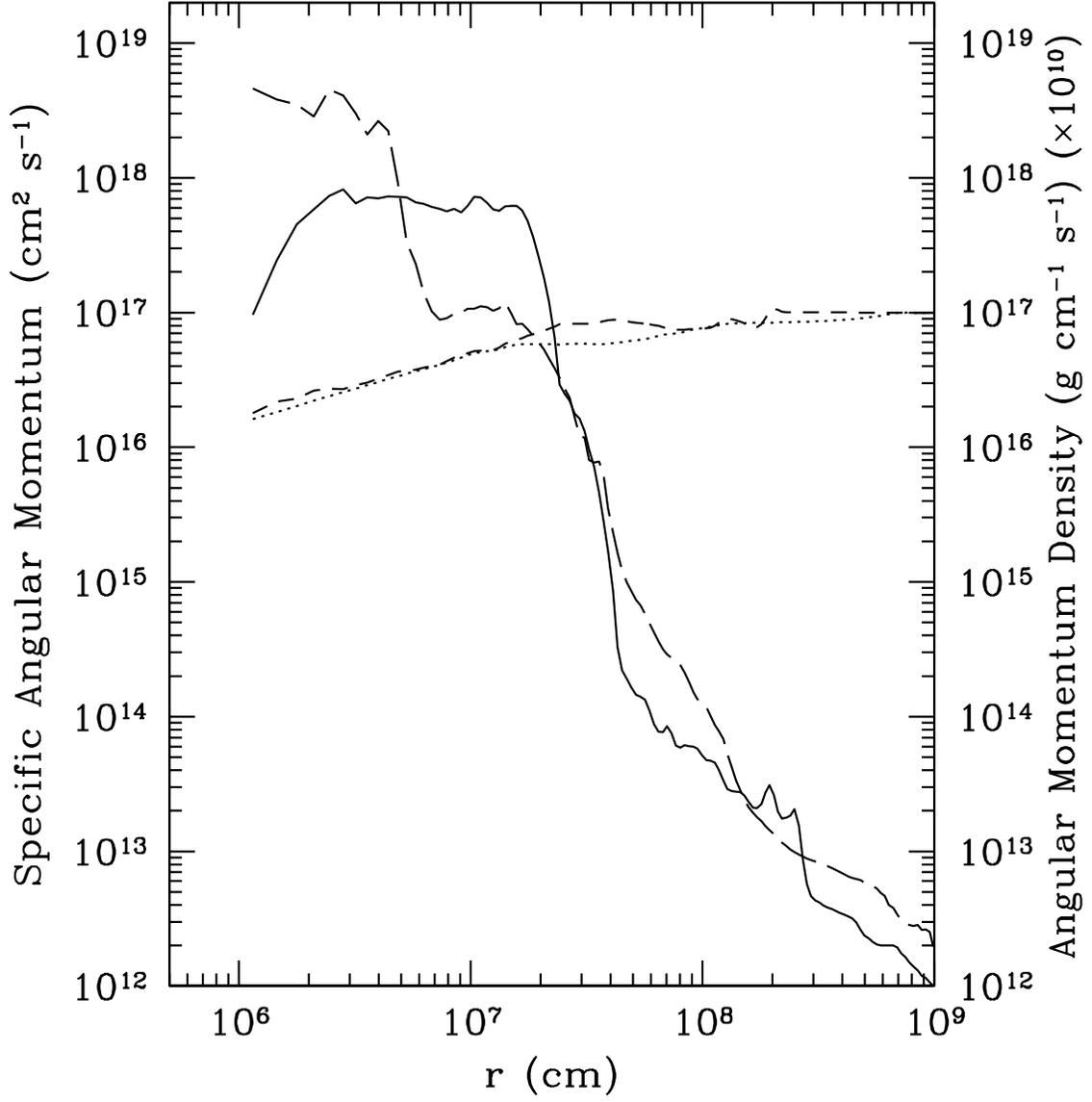}
\caption{
Profiles of angular momentum density (g cm$^{-1}$ s$^{-1}$)
for model 0 (long-dashed line) and model 9 (solid line)
on the equatorial plane
at $t$ = 2.2 s.
Profiles of specific angular momentum (cm$^2$ s$^{-1}$) 
are also shown
for model 0 (dotted line) and model 9 (short-dashed line). 
 \label{fig12}}
\end{figure}

In this study, we found that deposited energy due to neutrino pair
annihilation are too small to explain the explosion energy of a 
hypernova and a GRB (Fig.~\ref{fig9}). Even though the deposited
energy by electron-type neutrino capture on free nucleons can
be comparable to the explosion energy of a hypernova 
in model 9 (Fig.~\ref{fig9}),
the deposition region is the high density region (Fig.~\ref{fig7})
where cooling effect dominates the heating effect (Fig.~\ref{fig5}). 
In particular, no jet was found in the
numerical simulations of model 0 (Fig.~\ref{fig1}). The energetics 
of this system can be understood from Figs.~\ref{fig2},~\ref{fig6},
and~\ref{fig9}. The released gravitational energy by collapse is
the energy source of accreted energy and neutrinos. From
Figs.~\ref{fig2} and~\ref{fig6}, we can understand that the kinetic
energy and thermal energy share the released gravitational energy
almost equally, then almost thermal energy was extracted in the form of
neutrinos. The total energies of accreted energy and neutrinos are of
the order of 10$^{52}$ erg, then the emitted neutrino energy was
deposited into matter through weak interactions. Its efficiency is
less than 1$\%$ for neutrino pair annihilation, 
as can be seen in Figs.~\ref{fig6} and~\ref{fig9}. 
As for the efficiency of electron-type neutrino capture
it amounts to $\sim 10 - 20 \%$. This means that the inner most region
of the accretion disk becomes to optically thick against neutrinos
and $\sim 10 - 20 \%$ of neutrinos are absorbed.

The deposited energy by neutrino pair annihilation
is of the order of ($10^{49} - 10^{50}$  erg, which
is much smaller than the explosion energy of a hypernova and a GRB.
In order to enhance the deposited energy by neutrino pair annihilation, 
there will be two ways. 
One is to enhance the released gravitational energy, the
other is to enhance the efficiency of energy deposition.
The former corresponds to enhance the mass accretion rate, which will
be realized if effective angular momentum transfer is realized. 
From Figs.~\ref{fig2},~\ref{fig8}, and~\ref{fig12}, it was inferred
that magnetic fields seem to work efficiently so that high mass accretion
rate is realized. Of course, the mass accretion rate also depends on  
the distribution of initial angular momentum. We should investigate
these effects further in the future. As for the efficiency of energy
deposition, it will be enhanced when the general relativistic effects
are taken into account. This is because neutrinos are trapped around
the black hole, so that the possibilities of neutrino pair annihilation
and neutrino capture become enhanced. This effect is investigated in
detail by using a steady solution of an accretion disk
in the forth-coming paper. Of
course, we are planning to include this effect in our numerical code
in the future.

Although we believe that our conclusion on the energetics mentioned above
will be unchanged, we have to improve our treatment on the neutrino
heating for further study. In this study, we took the optical thin
limit to estimate the neutrino heating rate.
This will be
justified by Figs.~\ref{fig6} and~\ref{fig9}. 
However, for further study, we have to investigate the cases
in which mass accretion rate is higher than in this study to achieve
energetic explosion enough to explain the explosion energies of a
hypernova and a GRB. 
In fact, we consider that
the optically thin limit breaks down even the models
in this study at the highest density region. 
This is estimated as follows:
The cross section of $\nu_e$ and $\bar{\nu}_e$ captures on free nucleons
is given by $\sigma \sim \sigma_0 (\epsilon_{\nu}
/ m_e c^2)^2$ where $\sigma_0 = 1.76 \times 10^{-44}$ cm$^2$.
Since the highest density in the accretion disk is of the order of
$10^{12}$ g cm$^{-3}$ at $r \sim 10^{6}$ cm (the scale hight is also
of the order of $10^6$cm)
and typical energy of neutrinos are of the order of 10 MeV
(see Figs.~\ref{fig3} and~\ref{fig10}), the
optical depth at this region is
$\tau = \sigma (\rho /m_p) L \sim 4.2
(\epsilon_{\nu}/10 {\rm MeV})^2 (\rho / 10^{12} {\rm g \; cm ^{-3}}) (L /
10^6 {\rm cm}) $. Thus, at the highest density region, the optically thin
limit must break down. This picture is also confirmed in Fig.~\ref{fig8}(b).
In Fig.~\ref{fig8}(b), as stated in section~\ref{withmag}, the 
energy deposition rate due to pair captures on free nucleon sometimes 
becomes larger than the neutrino luminosity. This reflects that
the optically thin limit breaks down at that time. 
Although we believe that our conclusion on the energetics
will not be changed so much, we are planning to develop
the careful neutrino transfer code that includes emissions, absorptions,
and scattering of neutrinos for further study. 
We also note four points that have to be improved 
for the treatments of neutrino heating. One is that we did not
take into account the light-crossing time of the system and
assumed that the system is almost steady
during the light-crossing time
when we estimate the neutrino heating
rate. From 
Figs.~\ref{fig5} and~\ref{fig7}, 
the neutrino cooling and
heating occur efficiently within several times of $10^7$cm. Thus the typical
light crossing time will be of the order of 1ms. For comparison,
the rotation period
at the inner most region is $\sim 6.3 \times 10^{-4}$ s. 
Since the system forms an accretion disk and the viscosity parameter
$\alpha$ is 0.1 at most (Fig.~\ref{fig11}), the system will be
treated steady at least ten times of the rotation period,
$6.3 \times 10^{-3}$ s. Thus the
treatment to neglect the light crossing time will be fairly
justified. Second is that we update the neutrino heating rate every
100 timesteps to save the CPU time. The inner most radius is set to
be $10^6$ cm, so the typical time step is estimated to be $
10^6 \times \Delta \theta / c$
s, where $c$ is the speed of light and $\Delta \theta = \pi/60$. 
Thus 100 time steps corresponds to
$ 1.74 \sim  10^{-4}$ s, which will be comparable to free-fall
timescale ($\tau_{\rm ff} = 1 / \sqrt{24 \pi G \rho}  \sim 4.5 \times 10^{-4}
(10^{12} {\rm g \; cm^{-3}}/ \rho  )^{1/2}$ s~\cite{woosley86}) 
and rotation period. Thus we believe that this treatment will
be fairly justified. Third, we have approximated that each neutrino
energy spectrum due to each emission process is monotonic, that
is, we assumed only neutrinos with average energy are emitted when
Eq.~(\ref{Eq:anni}) is carried out. However, the cross section of
neutrino pair annihilation is proportional to square of the total
energy in the center of mass, and that for electron-type neutrino
absorption on free nucleons is proportional to square of neutrino's
energy. Thus the contribution of neutrinos with high energy will enhance
the efficiency of neutrino heating.
These points will be
improved when we can include a careful neutrino transfer code in
future. The last point is related with the nuclear reactions. In
this study, the NSE was assumed for the region where
$T \ge 5 \times 10^9$ [K]. Thus, the reactions to maintain NSE
occur suddenly when the temperature becomes so high as to satisfy 
the criterion. However, in reality, NSE might break down at
low density region, where cooling effect due to photo-disintegration
will be not so strong as in this simulation.
It should be also noted that
the photo-dissociation from He into nucleons is strong cooling effect
and absorb thermal energy when this reaction is switched on. 
Thus the thermal energy suddenly absorbed by
nuclear reactions. That is seen as the discontinuities of temperature
in Figs.~\ref{fig3} and~\ref{fig10}. 
Although we believe these discontinuities
do not change our conclusion on the energetics mentioned above
(because much more neutrinos come from the inner region;
Fig.~\ref{fig5}), 
the profile of temperature will be solved smoothly when
we use a nuclear reaction network instead of using the NSE
relation. Since the emissivity of neutrinos depends very sensitively
on the temperature~\citep{bethe90,herant92,lee06}, estimation of temperature
should be treated carefully. 
We are planning to check the dependence of temperature
on the nuclear reaction network and several EOS in the forth coming paper.

Finally, we discuss the detectability of neutrinos from
collapsars. Since the event rate is much smaller than the normal
core-collapse supernova, the chance probability to detect neutrino
signals from a collapsar will be very small. However, if it occurs
nearby our galaxy, the neutrino signal from a collapsar will be
distinguished from normal core-collapse supernova. As for the normal
core-collapse supernovae, the time evolution of the luminosity of
neutrino of each flavor is determined firmly by the binding energy of a
neutron star and opacity of neutron star against neutrinos. On the
other hand, in the case of a collapsar, the time evolution of neutrino
luminosity will depend on the time evolution of mass accretion rate,
which in turn should depend on the initial distribution of angular
momentum and magnetic fields. Thus there should be much varieties of
time evolution for the luminosity of neutrinos in the case of
collapsars. Also, in the case of collapsars, the dominant process to
generate neutrinos is pair captures on free nucleons (see
Fig.~\ref{fig6}), so in the case of a collapsar, the electron-type
neutrinos will be much more produced compared with other flavors. This
is in contrast with the normal core-collapse supernovae(e.g.
Buras et al. 2006, and see references therein). Of course,
we have to take vacuum and matter oscillation effects into account to
estimate the spectrum of neutrinos from a collapsar precisely. In
particular, in the case of a collapsar, the density distribution is
far from spherically symmetric, so we have to be careful about
viewing angle to estimate the matter oscillation effect.  
It is true that 
the event rate of collapsars is smaller than normal core collapse
supernovae, but the released gravitational energy can be larger
if considerable amount of mass of the progenitor falls into the
central black hole~\cite{nagataki02}.
Thus we consider that there will be also a possibility to detect
a neutrino background from collapsars.

\subsection{Effects of Magnetic Fields}\label{b-fields-dis}  

\begin{figure}
\epsscale{1.0}
\plotone{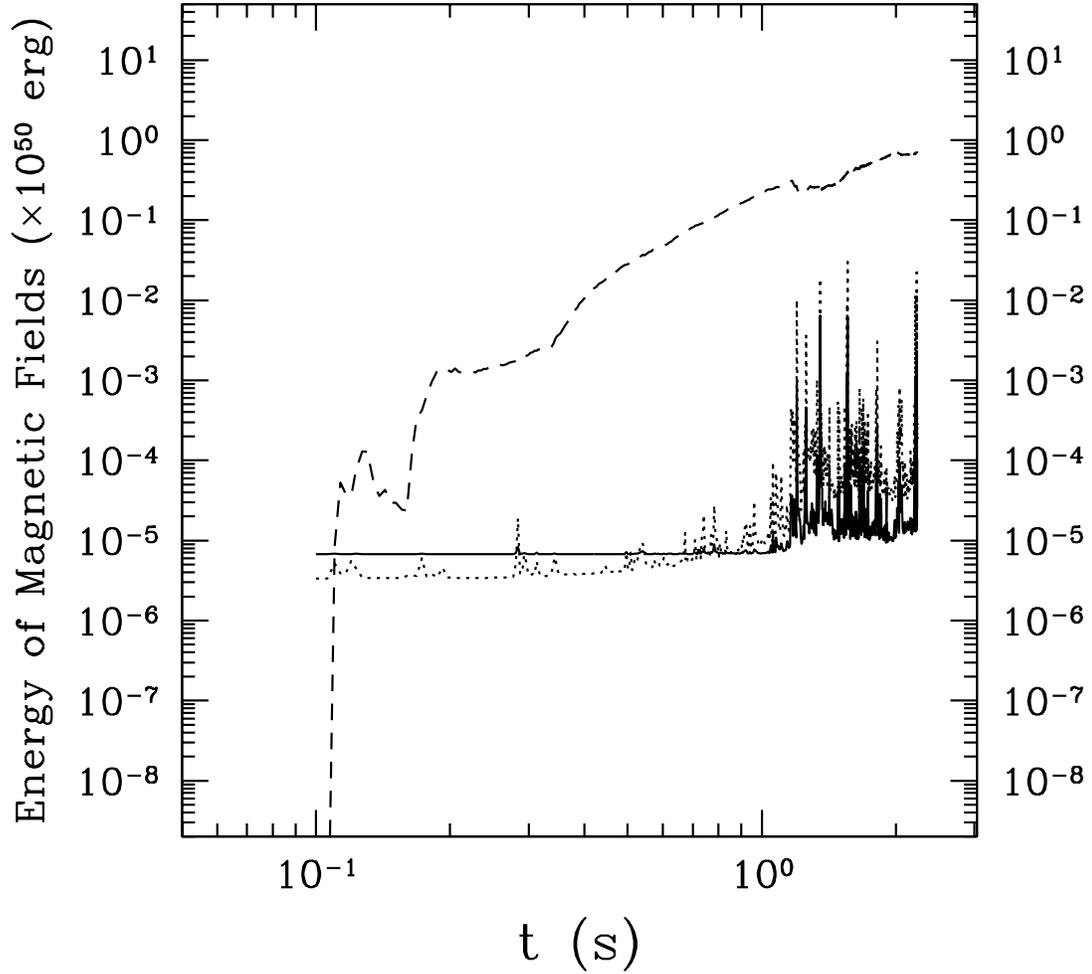}
\caption{Evolution of total energy of magnetic fields ($\times 10^{50}$ erg
) for the case of model 9. Dotted, solid, and dashed lines represent
energy in the form of $B_r$, $B_{\theta}$, and $B_{\phi}$, respectively.
Note that 1-D simulation of the spherical collapse of the progenitor
is done until $t=$0.1 s (see Section 2.2), so $B_{\phi}$ is set 
to be 0 until $t=0.1$ s.
 \label{fig13}}
\end{figure}

\begin{figure}
\epsscale{1.0}
\plotone{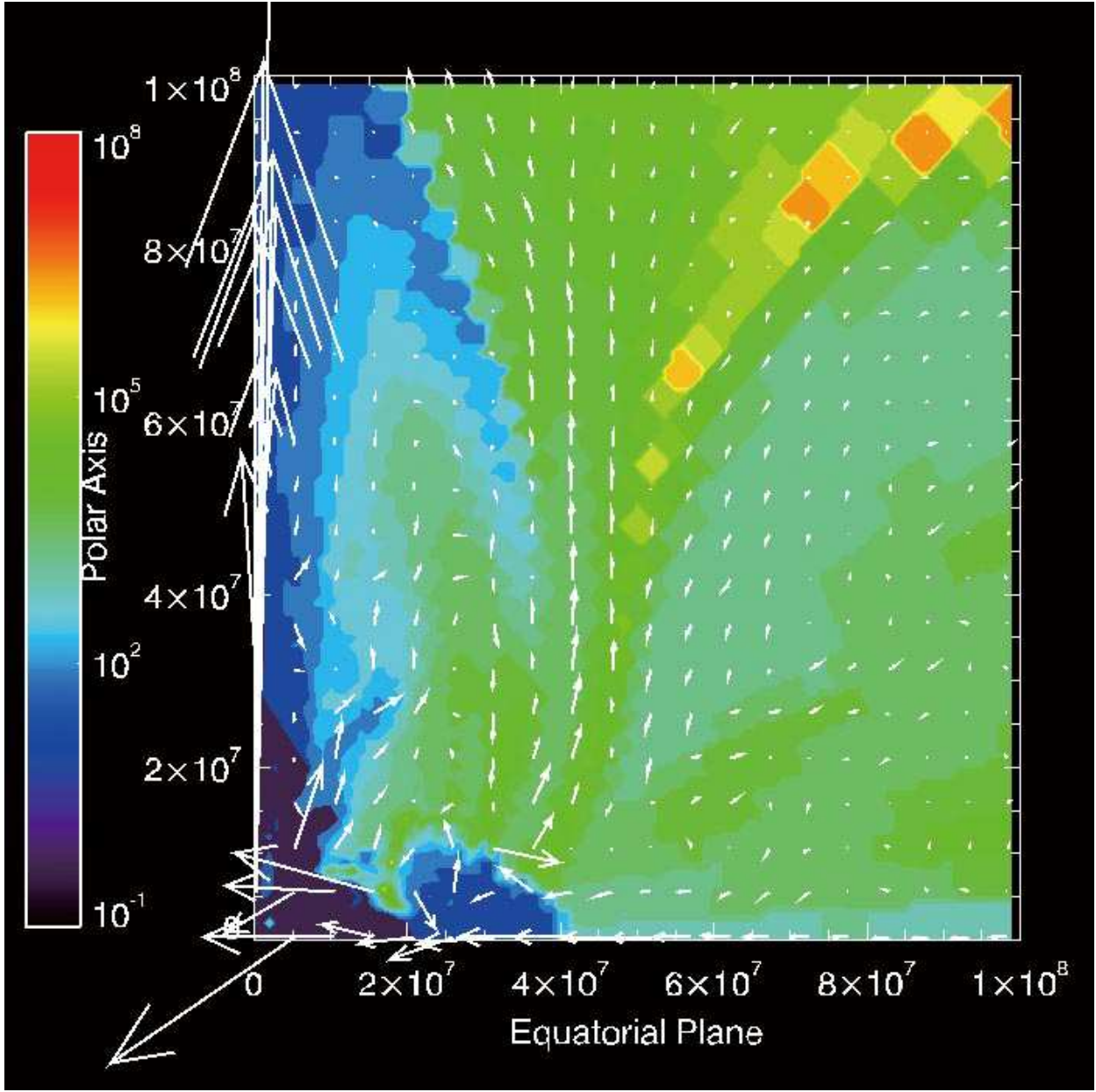}
\caption{Contour of plasma beta (=$p_{\rm gas+radiation}/p_{\rm mag}$)
with magnetic fields ($B_{r}$ and $B_{\theta}$) for model 9
at $t$ = 2.2 s. The color represents the plasma beta
in logarithmic scale ($10^{-1}-10^{8}$). 
 \label{fig14}}
\end{figure}

\begin{figure}
\epsscale{1.0}
\plotone{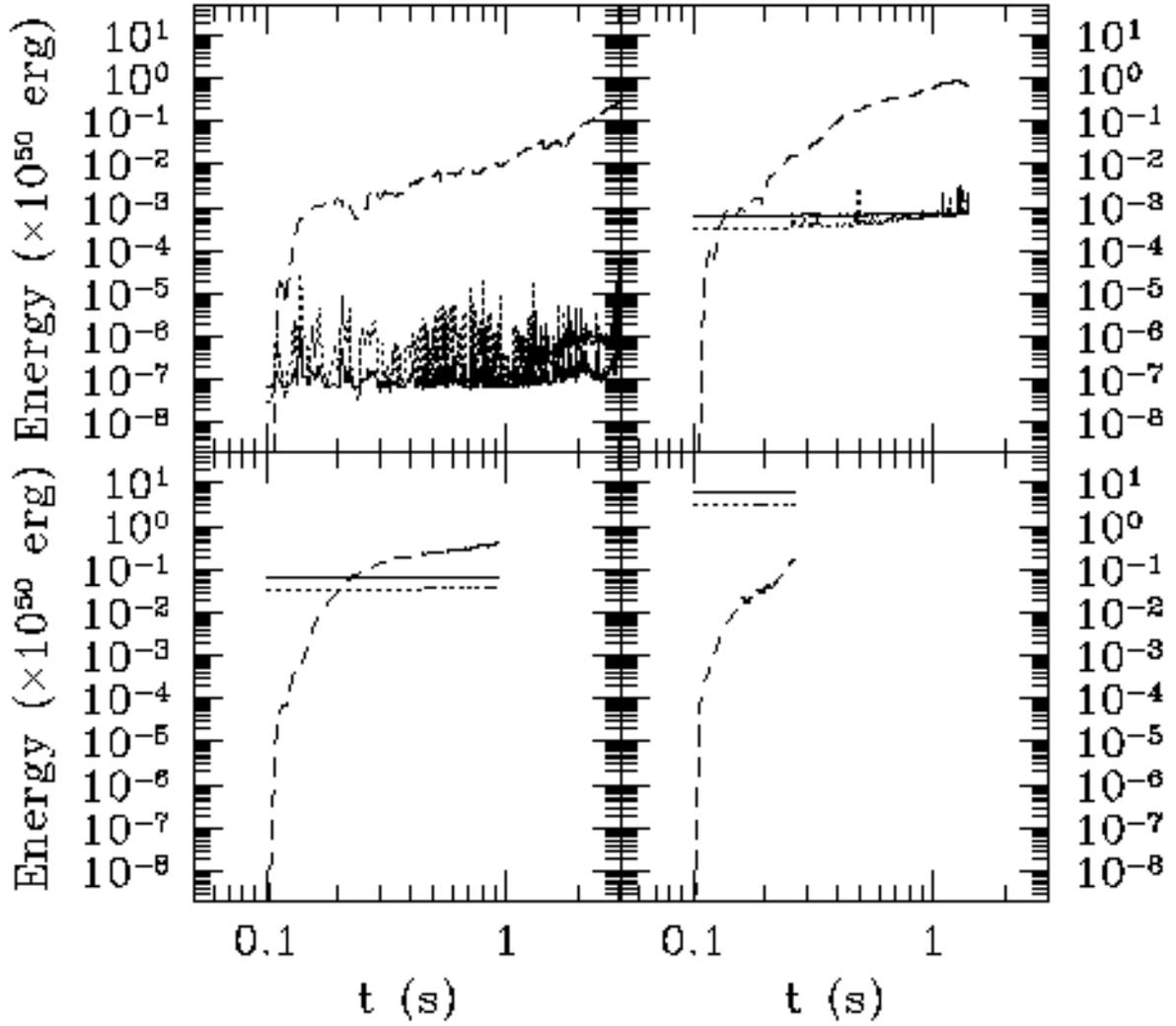}
\caption{Same as Fig. 13, but for model 8 (top-left panel), model 10
(top-right panel), model 11 (bottom-left panel), and model 12 (bottom
 right panel).
 \label{fig15}}
\end{figure}

\begin{figure}
\epsscale{1.0}
\plottwo{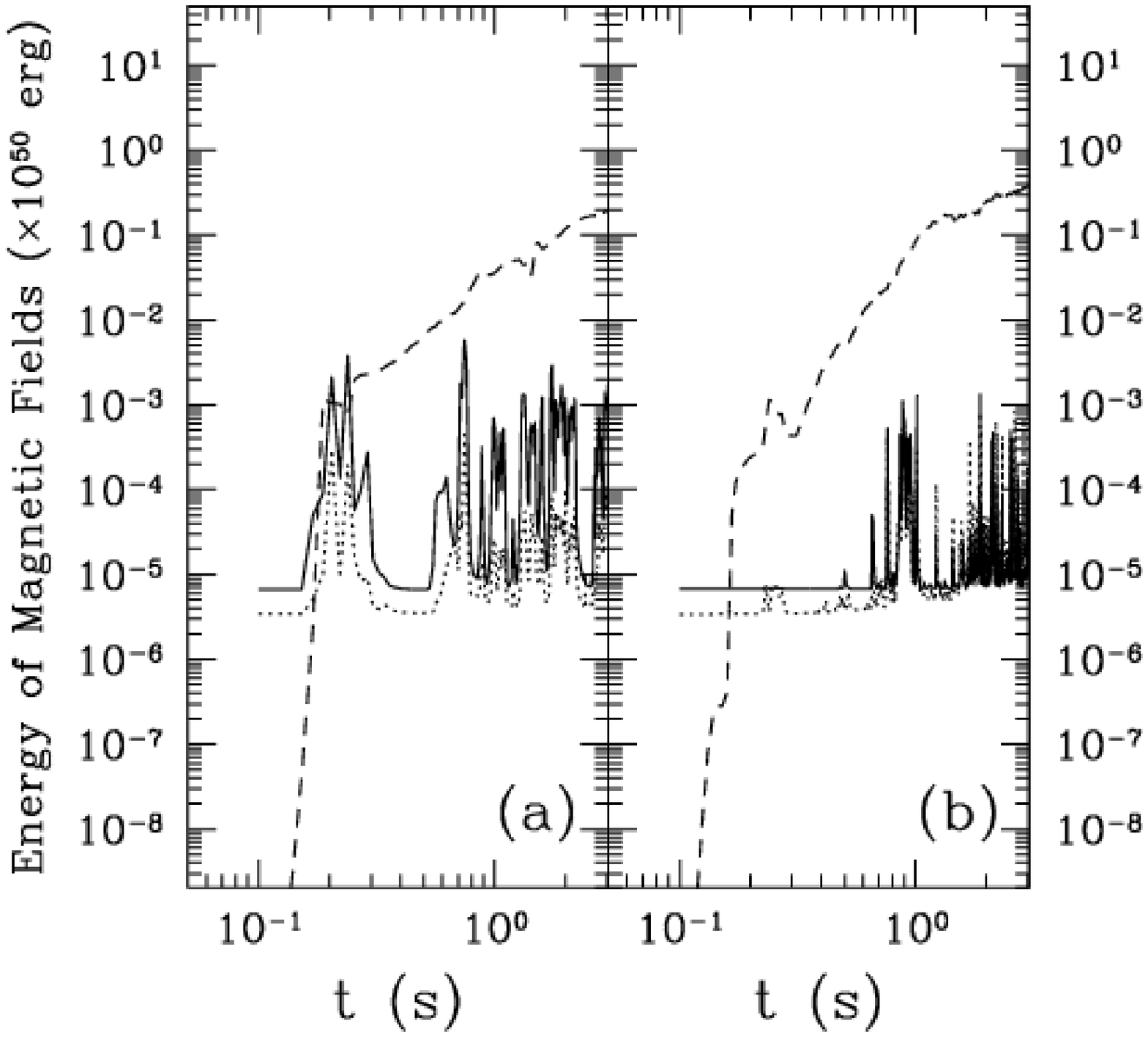}{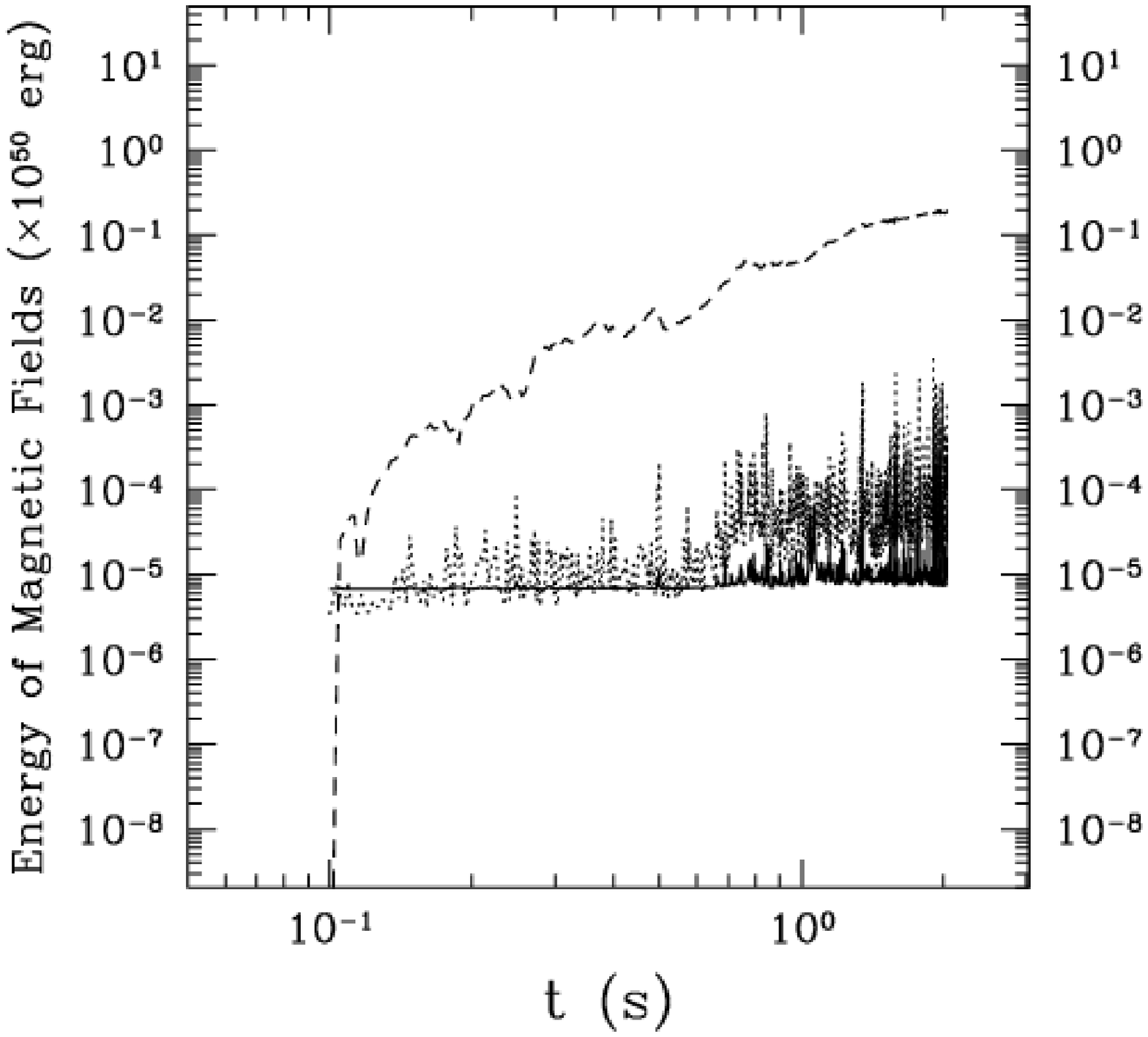}
\caption{Left panel: Same as Fig. 13, but for 150($r$)$\times$5($\theta$) grid
points (a) and 150($r$)$\times$20($\theta$) grid points (b). Right
 panel: Same as Fig. 13, but for 300($r$)$\times$60($\theta$) grid
points. In this simulation, neutrino anti-neutrino pair annihilation
effect is not included to save CPU time. Simulation region is 
(10$^6$cm $\le$ $r$ $\le$ $10^9$ cm; 0 $\le$ $\theta$ $\le$ $90$). 
 \label{fig16}}
\end{figure}

\begin{figure}
\epsscale{1.0}
\plotone{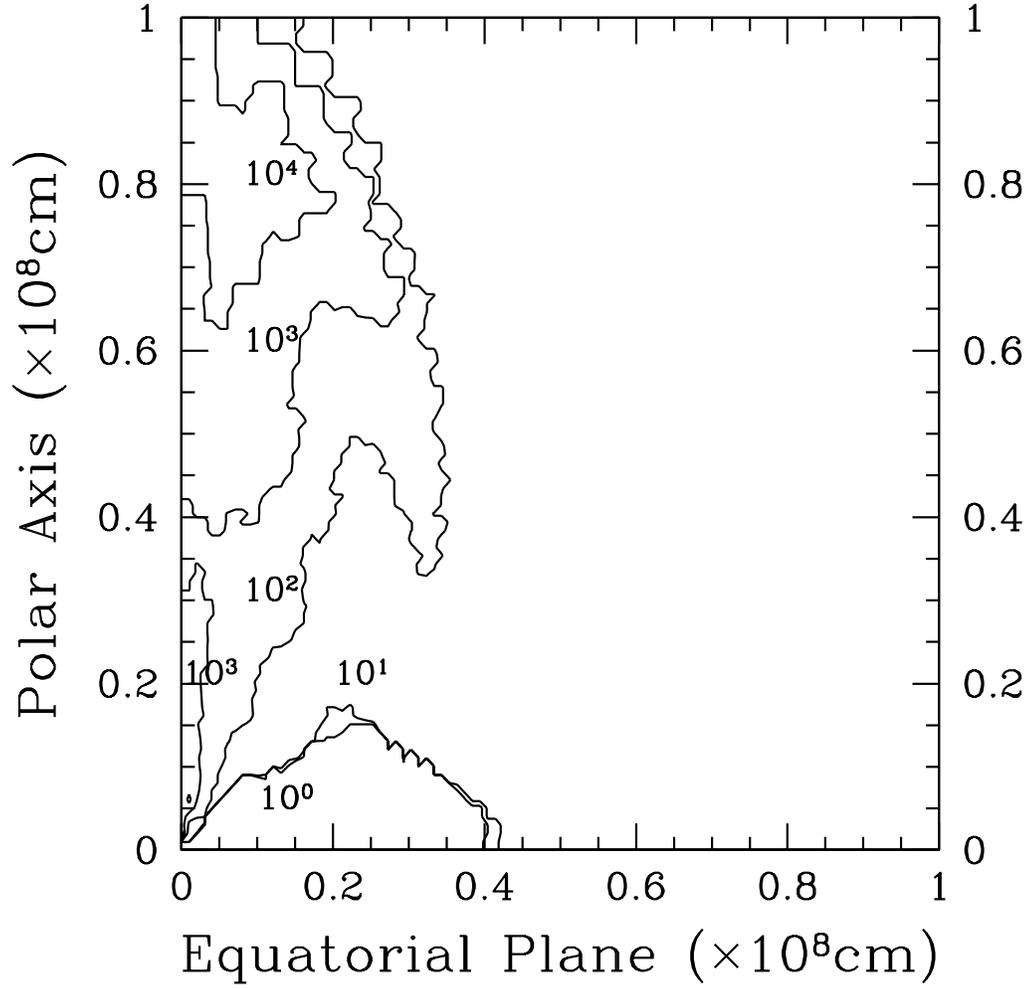}
\caption{
Contour of entropy per baryon in units of $k_b$ at $t=2.2$s for model 9.
The range of the contour is from 1 to $10^5$.
 \label{fig17}}
\end{figure}

\begin{figure}
\epsscale{1.0}
\plotone{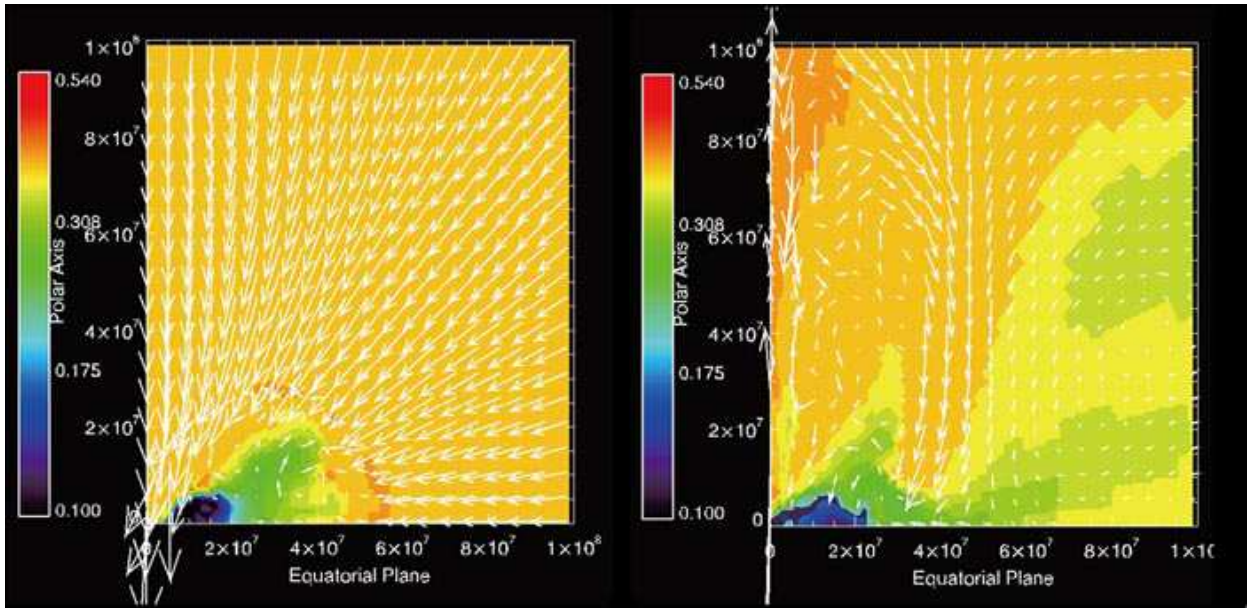}
\caption{
Contour of electron fraction with velocity fields for
for model 0 at $t$ = 2.2 s (left panel) and model 9 at $t$ = 2.2 s
(right panel). The color represents the electron fraction
on a linear scale (0.1-0.540). \label{fig18}}
\end{figure}

We have seen that the mass accretion rate seems to be enhanced in
model 9, compared with model 0 (Fig.~\ref{fig2}),
which enhances the luminosity of neutrinos
(Fig.~\ref{fig6}) and energy deposition rate due to weak interactions
(Fig.~\ref{fig8}). This seems to be because magnetic viscosity is
effective at inner most region (Fig.~\ref{fig11} and Fig.~\ref{fig12})
and multi-dimensional outflow (Fig.~\ref{fig1}) carries angular momentum
outward. 
Thus amplification of magnetic fields is
important not only for launching a jet by magnetic pressure, but also
for enhancing mass accretion rate and energy deposition rate through
weak interactions. 
In our study, we assume the axisymmetry of the system to save CPU
time. In this case, the field built up by the effect of
magnetorotational instability (MRI) decays due to Cowling's
anti-dynamo theorem~\cite{shercliff65}. However, the plasma beta
becomes lower than unity in the jet region (Fig.~\ref{fig14}), which
is embodied by the amplification of $B_\phi$ fields (Figs.~\ref{fig13}
and~\ref{fig15}). Thus we consider that $B_\phi$ field is not amplified 
by MRI effects, but by winding-up of poloidal fields due to 
differential rotation. 
The typical
timescale of winding-up at the inner most region will be (see
Figs.~\ref{fig3} and~\ref{fig10}) 
\begin{eqnarray}
\tau_{\rm wind} \sim 2 \pi \frac{d \ln r}{d \Omega} \sim \frac{2 \pi
\ln 10}{10^4} \sim 1.45 \times 10^{-3} \;\;\; \rm s.
\label{dis:b-1} 
\end{eqnarray}
This timescale will correspond to the steep rising of energy of
$B_\phi$ around $t=0.1$ s in Figs.~\ref{fig13} and~\ref{fig15}. 
After the steep growth, when the strength of $B_{\phi}$ becomes 
comparable to the 
poloidal component, the growth rate declines since $B_{\phi}$
grows by winding the 'weak' poloidal component~\cite{takiwaki04}.
The final time of the simulation
is determined when the Alfv\'{e}n speed reaches to the order of the
speed of light. This is understood as follows: 
Alfv\'{e}n crossing time in the $\theta$-direction at the innermost region
becomes $r \Delta \theta  /v_{\rm A} = 10^6 \times (\pi/60) / c = 1.74 
\times 10^{-6} $s.
Since the total timestep is several times of $10^6$, the final time is
estimated to be several time, which is consistent with our results. 
The Alfv\'{e}n speed is estimated to be
\begin{eqnarray}
v_{\rm A} = \frac{B}{\sqrt{4 \pi \rho}} &&\sim 2.82 \times 10^{8}
\left( \frac{B}{10^{15} \rm G} \right) \left(  \frac{ 10^{12}
\rm g \; cm^{-3}}{\rho} \right)^{1/2} \;\;\; \rm cm  \;s^{-1} \\
          &&\sim 2.82 \times 10^{10}
\left( \frac{B}{10^{15} \rm G} \right) \left(  \frac{ 10^{8}
\rm g \; cm^{-3}}{\rho} \right)^{1/2} \;\;\; \rm cm \;s^{-1},
\label{dis:b-2} 
\end{eqnarray}
which means that the final time is determined not by the time when
the amplitude of $B_\phi$ reaches to $10^{15}$G around the equatorial
plane, but by the time when the amplitude of $B_\phi$ reaches to
$10^{15}$G at low density region, that is, around the polar region
where the jet is launched.

As for $B_r$ and $B_\theta$ fields, from Fig.~\ref{fig13}, some instabilities
seem to grow and saturate, which is similar to the behavior of 
MRI~\citep{hawley91,balbus98}. 
At present, we consider that these instabilities are MRI modes with 
a wavelength of maximum growth mode unresolved. However, we can not 
conclude that these
instabilities are really unresolved MRI modes. This is due to the
reason as follows:
The dispersion relation of the linear MRI modes 
is obtained analytically by assuming that the accretion disk is supported by rotation 
(that is, in Kepler motion).
On the other hand, as shown in Figs.~\ref{fig3} and~\ref{fig10}, the radial velocity
is non-zero in the accretion disks in this study. Moreover, the radial
flow speed is
super-slow magnetosonic (the speed of slow magnetosonic wave is almost same with
that of Alfv\'{e}n wave in this study: Fig.~\ref{fig11}). 
Since MRI is the instability of the slow-magnetosonic waves in a
magnetized and differentially rotation plasma, the dispersion relation may be changed
considerably for such a super-slow magnetosonic flow. However, there is no analytic
solution for such a flow at present, so we use the dispersion relation of MRI for
the discussion here.

Ignoring entropy gradients, the
condition for the instability of the slow-magnetosonic waves in a
magnetized, differentially rotation plasma is~\cite{balbus91}
\begin{eqnarray}
\frac{d \Omega^2}{d \ln r} + (\bf{k} \cdot \bf{v}_{\rm A})^2 < 0,
\label{dis:b-3} 
\end{eqnarray}
where $\bf{k}$ is the vector of the wave number.
The wavelength of maximum growth of the linear instability is
\begin{eqnarray}
\lambda_0 = \frac{2 \pi v_{\rm A}}{\Omega} \sim 1.77 \times 10^3 
\left(  \frac{10^4 \rm s^{-1}}{\Omega} \right)
\left( \frac{B}{10^{13} \rm G} \right) \left(  \frac{ 10^{12}
\rm g \; cm^{-3}}{\rho} \right)^{1/2} \;\;\; \rm cm.
\label{dis:b-4}
\end{eqnarray}
Since $\lambda_0$ is much smaller than the grid size of the inner most
region ($\Delta r = 3 \times 10^5$cm), the linear MRI mode of maximum growth is not
resolved in this study (note that the amplitudes of $B_r$ and
$B_\theta$ are of the order of $10^{13}$G and much smaller
than $B_\phi$; see Fig.~\ref{fig11}). 
However, MRI grows as long as Eq.~(\ref{dis:b-3}) holds. Since
the value of the first term of Eq.~(\ref{dis:b-3}) at the
inner most region is $\sim -10^8/ \ln 10 \sim -4.34 \times 10^7$, it is 
confirmed that
inner most region is unstable for MRI mode with the wave length longer
than $\sim 10^3 (B/10^{13} \rm G)(10^{12} \rm g \;
cm^{-3}/\rho)^{1/2}$cm. The characteristic growing timescale
is~\citep{balbus98,akiyama03,proga03}
\begin{eqnarray}
\tau_{\rm MRI} \sim  2 \pi  
\left| \frac{d \Omega^2}{ d \ln r} \right|^{-1/2} \sim
6.74 \times 10^{-4} \;\;\; \rm s, 
\label{dis:b-5}
\end{eqnarray}
which is seen in model 9 (Fig.~\ref{fig13}) and model 8
(Fig.~\ref{fig15}). The saturation level of $B_r$ seems to be slightly
higher than that of $B_{\theta}$, which is similar to the results of the local
simulations of MRI~\cite{sano04}. Also, as shown in Fig.~\ref{fig16},
the growth rate of $B_{r}$ and $B_{\theta}$ components
depends on the grid resolution, while the growth rate of $B_{\phi}$ does not
depend on the grid resolution. In fact, the growth rate becomes smaller
for a coarse mesh case (Fig.~\ref{fig16}(a)).
This is similar to the picture that
$B_{r}$ and $B_{\theta}$ are amplified by MRI, while $B_{\phi}$ is
amplified by winding effect. The saturation level of $B_{r}$ and $B_{\theta}$
seems not to be sensitive to the grid resolution. 
As stated above, in order to prove firmly that
$B_{r}$ and $B_{\theta}$ are amplified by MRI-like instability, 
dispersion relation 
of linear growing modes for super-slow magnetosonic flow has to be obtained
analytically and the dispersion relation has to be reproduced by numerical simulations 
with finer grid resolution, which is outscope of this study.

As shown in Fig.~\ref{fig11}, the estimated viscosity parameter becomes larger
than $10^{-3}$ at almost all region in $r \le 4 \times 10^4$ cm.
Since the angular velocity becomes
larger than $10^3$ for $r \le 6 \times 10^6$ cm (Fig.~\ref{fig10}),
this viscosity becomes effective in a timescale of second, which is comparable 
to our simulations. Thus we think this viscosity drives high mass accretion rate
in the models with magnetic fields. 
However, as stated in section~\ref{withmag}, 
the angular momentum cannot be transfered to infinity
along the radial direction. 
As shown in Fig.~\ref{fig11}, the flow is super-Alfv\'{e}nic except for the inner 
most region. Thus angular momentum cannot be conveyed outward by Alfv\'{e}n wave 
in the radial direction. 
At present, we consider that the outflow (including the jet)
in the polar direction in model 9 should bring the angular momentum from 
the inner most region (see Fig.~\ref{fig1} bottom). This picture is
similar to CDAF~\cite{narayan01}. 
We are planning to present further analysis of this feature
in the forth-coming paper.

Since the plasma beta can be of the order of
unity at the inner most region (Fig.~\ref{fig14})
and $B_\phi$ dominates $B_r$ and $B_\theta$ at the
region ($r \le 7-8 \times 10^7$cm), we can understand that the magnetic
pressure from $B_r$ 
and $B_\theta$ is much smaller than the radiation pressure (from photons,
electrons, and positrons) and degenerate pressure. Thus, the jet cannot
be launched by the effects of $B_r$ and $B_\theta$ only. 
The winding-up effect is
necessary to amplify $B_\phi$ field so that the magnetic pressure
becomes comparable to the radiation and degenerate pressure.
We saw that MRI-like instability seems to occur from the beginning of the
simulations (see model 8 in Fig.~\ref{fig15}). However, since
the saturation
level is not so high, the amplified energy of magnetic fields for
$B_r$ and $B_\theta$ cannot be seen well at first (see model 9 in
Fig.~\ref{fig13} and model 10 in Fig.~\ref{fig15}). This is because
at the early phase the steady
accretion disk is not formed and released gravitational energy is not
so much (see Fig.~\ref{fig2}). In model 11 and model 12, the initial
total energies of magnetic fields are so high that the effect of MRI-like
instability cannot be seen (Fig.~\ref{fig15}).

As stated above, the axisymmetry of the system is assumed in this
study. So
winding up effect only amplifies $m=0$ mode of $B_\phi$. 
Usually, it is pointed out that saturation level of the winding up
effects becomes lower when
three dimensional simulations are performed~\cite{hawley95}. Thus
$B_\phi$ may not be amplified as strong as $10^{15}$G. 
Also, at the present study, the timescale for $B_\phi$ fields at the 
inner most region to be amplified to $\sim 10^{15}$G depends on the initial
amplitude of the magnetic fields (see Figs.~\ref{fig13} and~\ref{fig15}).
However, if three dimensional simulations are performed,
instabilities due to MRI(-like) modes do not decay due to Cowling's
anti-dynamo theorem.
Also, the MRI(-like) modes with wave vectors whose $\phi$
components are non-zero, which amplify $B_\phi$
are included~\cite{masada06}.
If such
MRI(-like) modes that amplify $B_\phi$ are included, the dependence 
of the dynamics on the initial amplitude of the magnetic fields may be 
diluted. We have to perform three dimensional calculations to see
what happens in more realistic situations.

As stated above, the timestep becomes so small when the Alfv\'{e}n
crossing time becomes so small. Since this calculation is Newtonian,
the speed of light is not included in the basic
equations for macro physics (Eq.~(\ref{Eq:basic0})-Eq.~(\ref{Eq:basic1})).
In fact, we found that the Alfv\'{e}n speed
becomes larger than the speed of light at some points by a factor of
two or so at the final stage of simulations. Thus it will be one of the
solutions to overcome this problem is to develop the special relativistic MHD
code, in which the Alfv\'{e}n speed is, of course, solved to be smaller than
the speed of light.

In this study, we considered the ideal MHD without dissipation (it is, 
of course, numerical viscosity is inevitably included due to finite 
gridding effects). When resistive heating is efficient,
considerable amount of energy in magnetic fields will be transfered to
thermal energy by ohmic-like dissipation and reconnection, which
will change the dynamics of collapsars so much. The problem is, however,
that the properties of resistivity of high density and high temperature matter
with strong magnetic fields are highly uncertain
(it is noted that artificial resistivity is included in Proga et
al. (2003) in order to account of the dissipation in a controlled way
instead of allowing numerical effects to dissipate magnetic fields
in an uncontrolled manner (see also Stone and Pringle 2001)).

Finally, we discuss the total explosion energy and bulk Lorentz factor of 
the jet. From Table~\ref{tab2}, we can see that there seems to be a tendency that
the mass of the jet becomes heavier when the initial amplitude of the magnetic
fields are stronger (model 10, model 11, and model 12), although this tendency
is not monotonic (model 8 and model 9).
This will reflect that the jet is launched earlier and the density is still 
higher for a case with stronger initial magnetic fields.
There seems to be also tendency that the energy of the jet becomes also larger 
when the initial amplitude of the magnetic fields is set to be stronger,
although this tendency is not also monotonic (model 8 and model 9).
This tendency is not so remarkable since the mass of the jet is heavier for a
stronger magnetic field case and the mass should have some amount of kinetic,
thermal, and magnetic field energies. As for the models 8 and 9, the
information
of initial condition might be lost considerably since it takes much time to launch jets. 
From these results, we can conclude that no GRB jet is realized
even if strong magnetic fields is assumed.
We consider that there will be a possibility that a GRB jet is realized
if we can perform numerical simulations for much longer physical time 
(say, $\sim 10-100$s). In such long timescale simulations, the energy
of $B_{\phi}$ fields should be much more than $10^{50}$ erg 
(see Figs.~\ref{fig13} and~\ref{fig15}). Also, 
the density in the jet may become lower along with time, because
considerable mass will falls into the black hole along the polar
axis. Thus the terminal bulk Lorentz factor may be enhanced at later phase. 
In order to achieve such a 
simulation, the special relativistic code will be helpful, 
as mentioned above.

\subsection{Prospect for Nucleosynthesis}\label{nucl-dis}  
It is radioactive nuclei, $\rm ^{56}Ni$ and its daughter nuclei,
$\rm ^{56}Co$, that brighten the supernova
remnant and determine its bolometric luminosity. $\rm ^{56}Ni$ is
considered to be synthesized through explosive nucleosynthesis
because its half-life is very short (5.9 days). Thus it is natural
to consider that explosive nucleosynthesis occurs in a hypernova
that is accompanied by a GRB. However, it is not
clearly known where the explosive nucleosynthesis occurs (e.g.
Fryer et al. 2006a, Fryer et al. 2006b).

Maeda et al. (2002) have done a numerical calculation of explosive
nucleosynthesis launching a jet by depositing thermal and kinetic
energy at the inner most region. They have shown that a 
mass of $\rm ^{56}Ni$ sufficient to explain the observation of hypernovae
($\sim 0.5 M_{\odot}$) can be synthesized around the jet region.
In their calculation, all the explosion energy was deposited initially.
Thus Nagataki et al. (2006) investigated the dependence of explosive
nucleosynthesis on the energy deposition rate. They have shown
that sufficient mass of $\rm ^{56}Ni$ can be synthesized as long as
all the explosion energy is deposited initially, while the synthesized
mass of $\rm ^{56}Ni$ is insufficient if the explosion energy is deposited
for 10 s (that is, the energy deposition rate is $10^{51}$ erg s$^{-1}$).
This is because matter
starts to move outward after the passage of the shock wave, and
almost all of the matter moves away from the central engine before
the injection of thermal energy (= $10^{52}$ erg) is completed, so
the amount of mass where complete burning to synthesize $\rm ^{56}Ni$
becomes little for such a long-duration explosion (see also Nagataki 
et al (2003)). 

On the other hand, it is pointed out the possibility that a substantial 
amount of $\rm ^{56}Ni$ is produced in the accretion disk and a part
of it is conveyed outward by the viscosity-driven
wind by some authors~\citep{macfadyen99,pruet03}. However, there is 
much uncertainty how much $\rm ^{56}Ni$
is ejected from the accretion disk. This problem depends sensitively
on the viscosity effects. Further investigation is required to
estimate how much $\rm ^{56}Ni$ is ejected.

In this study, we have found that mass of $\rm ^{56}Ni$ in the accretion
disk (see Fig.~\ref{fig4}) at the final stage of simulations is of the 
order of $10^{-3}M_{\odot}$ ($M_{\rm Ni}^{\rm tot}$ in Table~\ref{tab3}).
Thus if considerable fraction of the synthesized $\rm ^{56}Ni$ is ejected
without falling into the black hole, there will be a possibility to
supply sufficient amount of $\rm ^{56}Ni$ required to explain the
luminosity of a hypernova. However, in the present study, 
the ejected mass of $\rm ^{56}Ni$ was found to be only
$10^{-11}-10^{-6}M_{\odot}$ ($M_{\rm Ni}^{\rm esc}$ in Table~\ref{tab3}).
This is because the entropy per baryon in the jet is so high 
(Fig.~\ref{fig17}) that the light elements such as $n$, $p$, and He
dominate in the jet. Thus we could not show that sufficient mass of 
$\rm ^{56}Ni$ was ejected from the accretion disk in the present study.
There will be two possibilities to extract enough amount of
$\rm ^{56}Ni$ from the accretion disk. One is that $\rm ^{56}Ni$
might be extracted efficiently from the accretion disk at later phase.
Since typical temperature of the accretion disk will be lower when
the mass of the central black hole becomes larger, the entropy per baryon
in the jet will be decreased~\cite{nagataki02}. 
This feature should suggest that $\rm ^{56}Ni$
dominates in the jet component at the late phase. 
The other one is that some kinds of viscosities might work
to convey the matter in the accretion disk outward efficiently.
In MacFadyen and Woosley (1999), they included $\alpha$-viscosity and
showed that considerable amount can be conveyed.
Since $\alpha$ viscosity is not included in this study, such a feature
was not seen. However, when three dimensional simulations with higher
resolution than in this study are done, much more modes of magnetic
fields should be resolved, and some of them might be responsible for 
viscosities and work like $\alpha$-viscosity.

We discuss the possibility to synthesize heavy elements in collapsars.
We have shown that neutron-rich matter with high entropy per baryon
is ejected
along the jet axis (Figs.~\ref{fig17} and~\ref{fig18}). This is because
electron capture dominates positron capture in the accretion
disk (Fig.~\ref{fig6}). Thus there is a possibility that r-process
nucleosynthesis occurs in the
jet~\citep{nagataki97,nagataki00,nagataki01,nagataki01b,wanajo02,suzuki05,fujimoto06}.
Moreover, we found that mass element with high $Y_e$ ($\sim 0.522$) 
appears around the polar axis near the black hole. This is because
$\nu_e$ capture dominates $\bar{\nu}_e$ capture at the region.
This is because flux of $\nu_e$ from the accretion disk
(note that $\nu_e$ comes from electron capture)
is sufficiently large to enhance $\rm Y_e$ at the region.
In such a high $\rm Y_e$ region, there will be a possibility
that $r/p-$process nucleosynthesis occurs~\cite{wanajo06}. 
We are planning to perform
such a numerical simulations in the very near future. 
Also, there is a possibility that much neutrons are ejected
from the jet since the entropy per baryon amounts to the order of
$10^4$. If so, there may be a possibility that signals from the
neutron decays may be observed as the delayed bump of the light
curve of the afterglow~\cite{kulkarni05} or gamma-rays~\cite{razzaque06}.
Also, GeV emission may be observed by proton-neutron inelastic
scattering~\citep{meszaros00,rossi06,razzaque06b} 
Finally, note that a careful treatment of neutrino
transfer is important for the r- and r/p-process nucleosynthesis.
As stated in section~\ref{neutrino-dis}, the optically thin limit
may break down at the high density region. When the matter becomes
to be opaque to neutrinos, the neutrinos becomes to be trapped
and degenerate. In the high density limit, the chemical equilibrium
is achieved as $\mu_e + \mu_p = \mu_n + \mu_{\nu_e}$ where 
$\mu_e$, $\mu_p$, $\mu_n$, and $\mu_{\nu_e}$ are chemical potentials
of electrons, protons, neutrons, and electron-type neutrinos.  
When the chemical potential of electron-type neutrinos is not negligible,
the electron capture do not proceed further and electron fraction does
not decrease so much~\cite{sato75}, which should be crucial to the
r- and r/p-process nucleosynthesis.

\subsection{General Relativistic Effects}\label{GR-effect} 

In this section, we discuss general relativistic effects that we are
planning to include in
our magnetohydrodynamic simulation code.

First, we still believe that effects of energy deposition due to weak
interactions (especially, neutrino anti-neutrino pair annihilation) can be a key process
as the central engine of
GRBs. In fact, the temperature of the accretion disk becomes higher
especially for a Kerr
black hole since the gravitational potential becomes deeper and much
gravitational energy
is released at the inner most region and the radius of the inner most
stable orbit becomes
smaller for a Kerr black hole~\citep{popham99,macfadyen99}. This effect
will enhance the luminosity of the neutrinos from the accretion disk
(Popham et al. 1999),
and the neutrino pair annihilation~\citep{asano00,miller03,kneller06,gu06}. 
In the vicinity of the black hole, most of neutrinos
and anti-neutrinos come
shadows due to the bending effects of the neutrino geodesics. Since
the shape and position of
the black hole shadow depend on the physical parameters of the black
hole (Bardeen 1973; Takahashi 2004, 2005), the effective area emitting the most of the
neutrino flux just outside
of the black hole shadows is also determined by the black hole
parameters. These effects
are included only when geodesics and disk structures are precisely
calculated.

Since according to our simulations the plasma beta (=
pgas+radiation/pmag) in the polar
region is lower than the plasma beta in other regions and the minimum
values of the plasma
beta is 0.193. This means that the polar region is magnetically
dominated. In such regions
where the magnetic energy is dominated such as a force-free field, the
extraction of
the rotational energy of the black hole is expected as Poynting flux
(Blandford \& Znajek
1977) and the negative energy (Blandford \& Payne 1982; Takahashi et
al. 1990).
Although these effects have been confirmed numerically by many 
authors~\citep{koide03,mizuno04b,komissarov05,hawley06,mckinney06},
there are many unsolved issues on the central engine of GRBs.
Especially, the neutrino radiation from the negative energy fluid in
the realistic
general relativistic accretion flow is one of the unsolved issue. The
past studies roughly estimate
the extraction energy due to the Blandford-Znajek process to become
non-negligible
amount to the explosion of the long GRSs (e.g. Lee et al. 2000a,
2000b; Di Matteo et al. 2002).

\subsection{Variety of Initial Conditions}\label{inicon} 

The properties of a progenitor of GRBs are still unknown. There are much 
uncertainties on mass loss rate, final mass,
angular momentum distribution, amplitude and configuration
of magnetic fields prior to collapse. These properties will depend
not only on initial
mass and metallicity of progenitors but also on the presence
(or absence) and properties of companion stars~\citep{maeder01,woosley06,macfadyen05}. 
Thus it is necessary to perform numerical simulations with many models of
progenitors to ensure the validity of the mechanism to launch a GRB jet
from progenitors. In particular, we consider that the initial
angular momentum distribution should be important since the mass accretion
rate depends on it very sensitively.

\subsection{Prospects for Improvements of Numerical
Scheme}\label{numerical} 

As shown in Eq.~(\ref{dis:b-4}), the critical wavelength of maximum growth
of the linear MRI instability is quite short, so adaptive mesh refinement
method (e.g. Norman 2005; Zhang and MacFadyen 2006; Morsony et al. 2006)
is inevitably required to resolve the critical
wavelength in a collapsar. Of course, much CPU time is required to perform
three dimensional calculation with adaptive mesh refinement method.
Moreover, when we try to include neutrino transfer code, three dimensions
are required additionally for the momentum space. However, people will be
able to realize such an expensive simulation in the near future with a help of
rapidly growing power of super-computers. Of course, we are planning to
devote ourselves to developing such a numerical code.  

\section{SUMMARY AND CONCLUSION}\label{conclusion}
We have performed two-dimensional magnetohydrodynamic
simulations by the ZEUS-2D code
to investigate the dynamics of a collapsar that generates a GRB jet,
taking account of realistic equation of state (contribution of electrons,
positrons, radiation, and ideal gas of nuclei), neutrino cooling and
heating processes, magnetic fields, and gravitational force from the
central black hole and self gravity. 

We have found that neutrino heating processes (neutrino and anti-neutrino
pair annihilation, and $\nu_e$ and $\bar{\nu}_e$ captures on free nucleons)
are not so efficient to launch a jet. We have found that
a jet is launched by magnetic fields
(in particular, $B_\phi$ fields that are amplified by the winding-up effect).
However, the ratio of total energy relative to the rest mass energy in the jet
at the final stage of simulations suggests that the bulk Lorentz factor
of the jet will not reach to as high as several hundred, so we conclude
the jet seen in this study will not be a GRB jet.
We also found that the mass accretion rate seems to be enhanced in the models
with magnetic fields. This might be because angular momentum 
is efficiently transfered by the viscosity due to the magnetic fields
and multi-dimensional flow.

Since GRB jets are not obtained in this study,
we consider that general relativistic effects, by which
the efficiency of energy deposition through weak interactions will be enhanced
and rotation energy of the black hole will be transfered to matter through the
magnetic fields, will be important to generate a GRB jet. Thus we are planning to
develop a general relativistic magnetohydrodynamics code in the very near future.
Also, the accretion disk with magnetic fields may still play an important
role to launch a GRB jet,
and it may be seen if we can perform numerical simulations for much longer
physical time (say, $\sim 10-100$ s).
To realize such a simulation, the special
relativistic code will be helpful because the Alfv\'{e}n velocity is limited to
the speed of light.

We have shown that considerable amount of $\rm ^{56}Ni$ is synthesized
in the accretion disk. Thus if some fraction of the synthesized $\rm ^{56}Ni$
is ejected without falling into the black hole, there will be a possibility 
for the accretion disk to
supply sufficient amount of $\rm ^{56}Ni$ required to explain the
luminosity of a hypernova. 
Also, we have shown that neutron-rich matter with high 
entropy per baryon is ejected along the rotation axis. This is because
electron capture dominates positron capture. Moreover, we found that
the electron fraction becomes larger than 0.5 around the polar axis near 
the black hole. This is because $\nu_e$ capture dominates $\bar{\nu}_e$ capture
at the region. Thus there will be a possibility
that $r$-process and $r/p-$process nucleosynthesis occur at these regions.
Finally, much neutrons will be ejected
from the jet, which suggests that signals from the
neutron decays may be observed as the delayed bump of the light
curve of the afterglow or gamma-rays.

\acknowledgments
S.N. is grateful to S. Akiyama, R. Blandford, S. Fujimoto, S. Inutsuka,
S. Mineshige, T. Sano, S. Yamada, T. Yamasaki
and M. Watanabe for useful discussion. 
The computation was carried out on NEC SX-5 and SX-8, SGI
Altix3700 BX2, and Compaq Alpha Server ES40 at Yukawa Institute for
Theoretical Physics, Kyoto University. 
This work is in part supported by a Grant-in-Aid for the 21st Century
COE ``Center for Diversity and Universality in Physics'' from the
Ministry of Education, Culture, Sports, Science and Technology of
Japan. S.N. is partially supported by Grants-in-Aid for Scientific
Research from the Ministry of Education, Culture, Sports, Science and
Technology of Japan through No. 16740134. R.T. is partially supported
by Japan Society for the Promotion of Science No. 1710519. T.T is
partially supported by Japan Society for the Promotion of Science.

\clearpage

\begin{table*}
\begin{center}
\begin{tabular}{lcccccccccccccccc}
\tableline
\tableline
Model & $ \|T/W  \|$ ($\%$) & $\| E_m / W \|$ ($\%$)  & $B_0$ (G) \\
\tableline
Model 0  & 1.3 & 0                    & 0            \\
Model 8  & 1.3 & $1.1 \times 10^{-8}$ & $10^{8}$     \\
Model 9  & 1.3 & $1.1 \times 10^{-6}$ & $10^{9}$     \\
Model 10 & 1.3 & $1.1 \times 10^{-4}$ & $10^{10}$    \\
Model 11 & 1.3 & $1.1 \times 10^{-2}$ & $10^{11}$    \\
Model 12 & 1.3 & 1.1                  & $10^{12}$    \\
\tableline
\end{tabular}
\tablenum{1}
\caption{
$\|T/W  \|$: Initial ratio of the rotational energy to the
gravitational energy. $\| E_m / W \|$: Initial ratio of the magnetic
energy. $B_0$ is the strength of the magnetic field in the
sphere ($r < 3.6 \times 10^9$cm). The digit in the name of each model
represents the power index of $B_0$. 
}\label{tab1}
\end{center}
\end{table*}

\begin{table*}
\begin{center}
\begin{tabular}{lcccccccccccccccc}
\tableline
\tableline
Model & $E_{\rm J}$ (erg) & $ M_{\rm J}$ ($M_{\odot}$) & 
$\Gamma^{f}$-1 & $E_{\rm J}^{\rm mag}/E_{\rm J}$ \\
\tableline
Model 8 ($\theta_{\rm J} = 5^{\circ} $)   & 6.97E+47 & 4.88E-6 &
7.98E-2  & 4.75E-2\\
Model 8 ($\theta_{\rm J} = 10^{\circ} $)  & 7.40E+47 & 5.39E-6 &
7.67E-2  & 5.36E-2\\
Model 8 ($\theta_{\rm J} = 15^{\circ} $)  & 7.66E+47 & 5.56E-6 &
7.70E-2  & 6.12E-2\\
Model 9 ($\theta_{\rm J} = 5^{\circ} $)   & 2.96E+47 & 2.42E-7 &
6.81E-1  & 4.71E-2\\
Model 9 ($\theta_{\rm J} = 10^{\circ} $)  & 4.27E+47 & 3.13E-7 &
7.60E-1  & 3.60E-1\\
Model 9 ($\theta_{\rm J} = 15^{\circ} $)  & 4.87E+47 & 4.38E-7 &
6.21E-1  & 3.45E-2\\
Model 10 ($\theta_{\rm J} = 5^{\circ} $)  & 8.17E+45 & 1.97E-8 &
2.32E-1  & 2.83E-1\\
Model 10 ($\theta_{\rm J} = 10^{\circ} $) & 9.08E+45 & 2.12E-8 &
2.39E-1  & 2.72E-1\\
Model 10 ($\theta_{\rm J} = 15^{\circ} $) & 1.32E+46 & 3.65E-8 &
2.03E-1  & 2.13E-1\\
Model 11 ($\theta_{\rm J} = 5^{\circ} $)  & 1.05E+48 & 1.16E-5 &
5.05E-2  & 2.74E-1\\
Model 11 ($\theta_{\rm J} = 10^{\circ} $) & 1.16E+48 & 1.24E-5 &
5.22E-2  & 2.56E-1\\
Model 11 ($\theta_{\rm J} = 15^{\circ} $) & 1.32E+48 & 1.36E-5 &
5.44E-2  & 2.33E-1\\
Model 12 ($\theta_{\rm J} = 5^{\circ} $)  & 1.45E+49 & 1.16E-4 &
6.99E-2  & 1.69E-1\\
Model 12 ($\theta_{\rm J} = 10^{\circ} $) & 1.77E+49 & 1.52E-4 &
6.50E-2  & 1.57E-1\\
Model 12 ($\theta_{\rm J} = 15^{\circ} $) & 2.39E+49 & 2.47E-4 &
5.42E-2  & 1.29E-1\\
\tableline
\end{tabular}
\tablenum{2}
\caption{
$E_{\rm J}$ (erg), $ M_{\rm J}$ ($M_{\odot}$), $\Gamma^{f}$, and
$E_{\rm J}^{\rm mag}/E_{\rm J}$ are mass, total energy, terminal bulk
Lorentz factor, and ratio of the magnetic energy relative to total
energy of the jet at the final stage of the simulations,
respectively. $\theta_{\rm J}$ is the assumed
opening angle of the jet. See section 3.3.1 for details. 
}\label{tab2}
\end{center}
\end{table*}

\begin{table*}
\begin{center}
\begin{tabular}{lcccccccccccccccc}
\tableline
\tableline
Model &  $ M_{\rm Ni}^{\rm esc}$
 ($M_{\odot}$) & $M_{\rm Ni}^{\rm tot}$ ($M_{\odot}$) \\
\tableline
Model 8  & 6.04E-10 & 3.79E-3   \\
Model 9  & 7.52E-11 & 5.92E-3   \\
Model 10 & 1.09E-7  & 3.95E-3   \\
Model 11 & 1.40E-6  & 1.61E-3   \\
Model 12 & 5.62E-7  & 1.12E-3   \\
\tableline
\end{tabular}
\tablenum{3}
\caption{
$ M_{\rm Ni}^{\rm esc}$ ($M_{\odot}$)
and $M_{\rm Ni}^{\rm tot}$ ($M_{\odot}$) represent the mass of 
$\rm ^{56}Ni$ in the regions where total energy (i.e. summation of kinetic
energy, thermal energy, and gravitational energy) is positive and
the total mass of $\rm ^{56}Ni$ in the whole simulated region
at the final stage of the simulations, respectively.
}\label{tab3}
\end{center}
\end{table*}


\end{document}